\documentclass[runningheads]{svmult}
\usepackage{makeidx}   
\usepackage{graphicx}  
\usepackage{subeqnar}  
\usepackage{multicol}  
\usepackage{physprbb}  
\usepackage{amssymb}
\usepackage{bookmath}
\makeindex             
\begin{document}
\title*{Vortex Core Structure and Dynamics \\ 
	\hspace*{10pt} in Layered Superconductors}
\toctitle{Vortex Core Structure and Dynamics in Layered Superconductors}
\titlerunning{Vortex Core Structure and Dynamics}
\author{M. Eschrig\inst{1,2}
        \and D. Rainer\inst{3}
        \and J. A. Sauls\inst{1}
}
\authorrunning{Matthias Eschrig et al.}
\institute{
	Department of Physics, Northwestern University, Evanston, IL 60208, USA
	\and 
	Materials Science Division, Argonne National Laboratory, Argonne, IL 60439, USA
	\and
	Physikalisches Institut, Universit\"at Bayreuth, D-95440 Bayreuth, Germany 
}
\maketitle 
\vspace{-70 mm}\noindent{\tiny Presented at the 2000 Workshop on 
                       {\sl Microscopic Structure and Dynamics of Vortices in 
		       Unconventional Superconductors and Superfluids}, 
		       held at the {\sl Max Planck Institute for the 
		       Physics of Complex Systems} in Dresden.}
\vspace*{65 mm}
\begin{abstract}
We investigate the equilibrium and nonequilibrium properties of the core region 
of vortices in layered superconductors. We discuss the electronic structure of 
singly and doubly quantized vortices for both s-wave and d-wave pairing symmetry. 
We consider the intermediate clean regime, where the vortex-core bound states 
are broadened into resonances with a width comparable to or larger than the 
quantized energy level spacing, and calculate the response of a vortex core to an 
{\em a.c.} electromagnetic field for vortices that are pinned to a metallic defect.
We concentrate on the case where the vortex motion is nonstationary and
can be treated by linear response theory.
The response of the order parameter, impurity self energy, 
induced fields and currents are obtained by a self-consistent calculation
of the distribution functions and the excitation spectrum. 
We then obtain the dynamical conductivity, spatially resolved in 
the region of the core, for external frequencies in the range, 
$0.1\Delta < \hbar\omega \lsim 3\Delta$. We also calculate 
the dynamically induced charge distribution in the vicinity
of the core. This charge density is related to the nonequilibrium 
response of the bound states and collective mode, and
dominates the electromagnetic response of the vortex core.
\end{abstract}

\section{Introduction}\label{Sec:Introduction}

Vortex motion is the principle mechanism for resistive losses
in type II superconductors.
Vortices also provide valuable information about the
nature of low lying excitations in the superconducting state.
In clean $s$-wave BCS superconductors the low-lying 
excitations in the core are the bound states of 
Caroli, de Gennes and Matricon \cite{car64}.
These excitations have superconducting as well as normal 
metallic properties. For example, these states are the 
source of circulating supercurrents in the equilibrium vortex core, 
and they are strongly coupled to the condensate by Andreev scattering 
\cite{bar69,rai96}. Furthermore, the response of the vortex core states 
to an electromagnetic field is generally very different
from that of normal electrons. 
However, in the dirty limit, $\hbar/\tau\gg\Delta$, the the Bardeen-Stephen 
model \cite{bar65} of a normal-metal spectrum with the
local Drude conductivity in the core
provides a reasonable description of the dissipative dynamics of the vortex core.
The opposite extreme is the ``superclean limit'',  $\hbar/\tau\ll\Delta^2/E_f$,
in which the quantization of the vortex-core bound states must be taken into account.
In this limit a single impurity and its interaction with the vortex core states 
must be considered. The a.c. electromagnetic response is then controlled 
by selection rules governing transition matrix elements for the quantized 
core levels and the level structure of the core states in the presence
of an impurity \cite{lar98,kou99,atk99}.
In the case of d-wave superconductors in the superclean limit, {\sl nodes}
in the spectrum of bound states lead to a finite dissipation from Landau
damping for $T\rightarrow 0$ \cite{kop97}.

The superclean limit is difficult to achieve even for short coherence 
length superconductors; weak disorder broadens the vortex 
core levels into a quasi-continuum. We investigate the intermediate-clean 
regime, $\Delta^2/E_f\ll\hbar/\tau\ll\Delta$, where the discrete
level structure of the vortex-core states is broadened and the selection 
rules are broken {due to strong overlap between the bound state wave functions}. 
However, the vortex core states remain well defined on the
scale of the superconducting gap, $\Delta $. In this regime we can 
take advantage
of the power of the quasiclassical theory of nonequilibrium 
superconductivity \cite{eil68,lar68,eli71,lar76,lar77}.

The energy required to maintain a net charge density
of order an elementary charge per particle within
a coherence volume (or coherence area in two dimensions) 
is much larger than the condensation energy.
Thus, charge accumulation in the vortex core is strongly suppressed.
In order to reduce the Coulomb energy associated with the charge 
accumulation an internal electrochemical potential, $\Phi(\vec{R};t)$,
develops in response to an external electric 
field. This potential produces an internal electric field, 
$\vec{E}^{\mbox{\tiny int}}(\vec{R};t)$, which is the same
order of magnitude as the external field. Even though the external field may
vary on a scale that is large compared to the coherence length, $\xi_0$,
the internal field 
develops on the coherence length scale. The source of the internal field is a
charge density that accumulates inhomogeneously over length scales of order
the coherence length. It is necessary to calculate the induced potential 
self consistently from the spatially varying order parameter,
spectral function and distribution function for
the electronic states in the vicinity of the vortex core.
An order of magnitude estimate shows that to produce an induced
field of the order of the external field, the
dynamically induced charge is of order $e\;(\Delta/E_f)(\delta v_{\omega}/\Delta)$,
where $\delta v_{\omega}\sim e{E}^{\mbox{\tiny ext}}/\xi_0\omega$ is 
the typical energy scale set by the strength of the external field.
This charge density accumulates predominantly in the vortex
core region and creates a dipolar field around the vortex core. 
For a pinned vortex the charge accumulates near the 
interface separating the metallic inclusion from the superconductor.

Disorder plays a central role in the dissipative dynamics of the
mixed state of type II superconductors. Impurities and defects are a source
of scattering that limits the mean free path of carriers, thus increasing
the resistivity. Defects also provide `pinning sites' that inhibit
vortex motion and suppresses the flux-flow resistivity.
However, for {\em a.c.} fields even pinned vortices are sources for dissipation.
The magnitude and frequency dependence of this dissipation depends on the electronic
structure and dynamics of the core states of the pinned vortex. In the analysis 
presented below we consider vortices in the presence of pinning centers.
We model a pinning center as a normal metallic inclusion 
which is coupled to the electronic
states of the superconductor through a highly 
transmitting interface. In this model the 
charge dynamics of the electronic states near
the interface between the pinning region and 
superconductor plays an important role in the 
electromagnetic response of the core.

In the next section we provide a short summary of the 
nonequilibrium quasiclassical equations, including the transport 
equations for the quasiparticle distribution and spectral functions,
constitutive equations for the order parameter, impurity self-energy 
and electromagnetic potentials. In section \ref{Sec:Electronic_Structure} 
we present calculations for the the electronic 
structure of vortices for superconductors with both $s$-wave and 
$d$-wave pairing symmetry. The results are based on self-consistent
calculations of the order parameter and impurity self energy. 
For $s$-wave superconductors, impurity scattering leads to inhomogeneous
broadening of the vortex core bound states, as well as bands of impurity 
states within the gap. In the case of $d$-wave pairing the core
states are further broadened by coupling between bound states
and the continuum states through impurity scattering. We also
discuss the structure of doubly quantized vortices and vortices
bound to mesoscopic size metallic inclusions. In the case
of the doubly quantized vortex there are two branches of 
zero-energy bound states centered at finite impact parameter 
from the vortex center. This leads to a unique signature of a
doubly quantized vortex: currents in the core circulate 
opposite to the supercurrents outside the core. 
Section \ref{Sec:Nonequilibrium_Response} summarizes calculations
of the vortex core dynamics for $s$-wave vortices in the presence
of impurity scattering. We describe the charge dynamics of
the vortex core for both pinned and unpinned vortices, and 
calculate the local a.c. conductivity that results from the coupled
dynamics of the order parameter collective mode and the quasiparticle
bound states in the vortex core. We discuss
energy transport by the core states and the absorption features in the 
conductivity spectrum, which we interpret in terms of absorption
within the bound-state band centered at the Fermi level
and resonant transitions involving the bound and continuum states.

\section{Nonequilibrium Transport Equations}\label{Sec:Transport_Theory}

The quasiclassical theory describes equilibrium and nonequilibrium
properties of superconductors on length scales that are large compared 
to microscopic scales (i.e. the lattice constant, Fermi wavelength, 
$k_f^{-1}$, Thomas-Fermi screening length, etc.) and
energies that are small compared to the atomic
scales (e.g. Fermi energy, $E_f$, plasma frequency, 
conduction band width, etc.). Thus, there are small dimensionless
parameters that define the limits of validity of the quasiclassical
theory. In particular, we require $k_f\xi_0\gg 1$, $k_B T_c/E_f\ll 1$
and $\hbar\omega\ll E_f$, where the {\it a.c.} frequencies of interest 
are typically of order $\Delta\sim T_c$, or smaller, and the 
length scales of interest are of order the coherence length, 
$\xi_0= \hbar v_f/2\pi k_BT_c$, or longer. Hereafter we use 
units in which $\hbar=k_B=1$, and adopt the sign convention 
$e =-|e|$ for the electron charge.

In quasiclassical theory quasiparticle wavepackets move along nearly straight, 
classical trajectories at the Fermi velocity. The classical dynamics of the
quasiparticle excitations is governed by semi-classical transport equations for
their phase-space distribution function.  The quantum mechanical degrees of 
freedom are the ``spin'' and ``particle-hole degree 
of freedom'', described by $4\times 4$ density matrices (Nambu matrices). 
The quantum dynamics is coupled to the classical dynamics of the 
quasiparticles in phase space through the matrix structure of the 
quasiclassical transport equations.

The nonequilibrium quasiclassical transport equations
\cite{eil68,lar68,eli71,lar76,lar77}
are formulated in terms of a quasiclassical Nambu-Keldysh propagator
$\check{g}(\vec{p}_f,\vec{R};\epsilon,t)$, which is a matrix in the 
combined Nambu-Keldysh space, and is a function of position $\vec{R}$, 
time $t$, energy $\epsilon$, and  momenta $\vec{p}_f$ on the Fermi 
surface.\footnote{In quasiclassical theory the description in terms of 
the variables ($\epsilon$,$\vp_f$,$\vR$) is related to the
phase-space description in $\vp-\vR$ space by a transformation,
$g(\vp_f,\epsilon;\vR,t)=f(\vp,\vR;t)$, with 
$\epsilon=\varepsilon(\vp,\vR;t)-\mu$ and $\hat{\vp}=\hat{\vp}_f$
\cite{ser83}.}
We denote Nambu-Keldysh matrices by a ``check'', and their 
4$\times$4 Nambu submatrices of {\em advanced} ($A$), {\em retarded} ($R$) 
and {Keldysh}-type ($K$) propagators by a ``hat''.  The Nambu-Keldysh 
matrices for the quasiclassical propagator and self-energy have the form,
\be\label{Nambu-Keldysh}
\check{g}=\left(\matrix{\hat{g}^R&\hat{g}^K\cr 0 &\hat{g}^A}\right)
\qquad ,\qquad
\check{\sigma}=
\left(\matrix{\hat{\sigma}^R&\hat{\sigma}^K\cr 0& \hat{\sigma}^A}\right)
\,,
\ee
where $\hat{g}^{R,A,K}$ are the retarded (R), advanced (A) and Keldysh (K) 
quasiclassical propagators, and similarly for the self-energy functions. 
Each of these components of $\check{g}$ and $\check{\sigma}$
are $4\times 4$ Nambu matrices in combined particle-hole-spin space.
For a review of the methods and an introduction to the notation we refer 
to Refs. \cite{lar86,ser83,rai95}.
In  the compact Nambu-Keldysh notation the transport equations and  the
normalization conditions read
\begin{equation}\label{transpqcl}
\bigl[(\epsilon+{e\over c}\vec{v}_f\cdot\vec{A})\check\tau_3-eZ_0\Phi \check{1}
-\check\Delta_{mf}-\check\nu_{mf}-\check\sigma_{i} \,,\check
g\bigr]_{\otimes} +i\vec{v}_f\cdot\grad \check g\\ = \ 0\enspace ,
\end{equation}
\begin{equation}\label{normalize}
\check g\otimes \check g=-\pi^2\check{1}\enspace ,
\end{equation}
where the commutator is 
$[\check A,\check B]_\otimes=\check A\otimes\check B-\check B\otimes\check A$, 
\begin{equation}
\check A \otimes \check B(\epsilon,t) =
e^{\frac{i}{2}(\partial_\epsilon^A \partial_t^B-\partial_t^A\partial_\epsilon^B)}
\check A(\epsilon,t)\check B(\epsilon,t)
\,.
\end{equation}
The vector potential, $\vec{A}(\vec R;t)$, includes 
$\vec{A}_0(\vec R)$ which generates the static magnetic field, 
$\vec{B}_0 (\vec R) =\grad \times\vec{A}_0(\vec R )$, as well as
the non-stationary vector potential describing the time-varying 
electromagnetic field; $\check\Delta_{mf}(\vec{p}_f,\vec{R};t)$ 
is the mean-field order parameter matrix, 
$\check\nu_{mf}(\vec{p}_f,\vec{R};t)$ describes diagonal 
mean fields due to quasiparticle interactions (Landau interactions),
and $\check\sigma_{i}(\vec{p}_f,\vec{R};\epsilon,t)$ is the impurity
self-energy. The electrochemical potential $\Phi (\vec{R};t)$ includes
the field generated by the induced charge density, $\rho(\vec{R};t)$. 
The coupling of quasiparticles to the external potential involves 
virtual high-energy processes, which result from polarization of the 
non-quasiparticle background. The interaction of quasiparticles with 
both the external potential $\Phi$ and the polarized background can be 
described by coupling to an effective potential $Z_0\Phi$ \cite{ser83}.
The high-energy renormalization factor $Z_0$ is defined below 
in Eq. (\ref{renormalization}). The coupling of the quasiparticle 
current to the vector potential in Eq. (\ref{transpqcl}) is given in 
terms of the quasiparticle Fermi velocity. No additional renormalization 
is needed to account for the effective coupling of the charge current to 
the vector potential because the renormalization by the non-quasiparticle 
background is accounted for by the effective potentials that 
determine the band structure, and therefore the quasiparticle Fermi 
velocity.

\subsection{Constitutive Equations}

Equations (\ref{transpqcl}-\ref{normalize}) 
must be supplemented by Maxwell's 
equations for the electromagnetic potentials,
and by self-consistency equations 
for the order parameter and the impurity self-energy.
We use the weak-coupling gap equation to describe 
the superconducting state, including unconventional 
pairing. The mean field self energies are then given by,
\begin{equation}\label{gapequation}
\hat\Delta^{R,A}_{mf}(\vec{p}_f,\vec{R};t)= N_f
\int_{-\epsilon_c}^{+\epsilon_c}{d\epsilon\over 4\pi i}
\big<
V(\vec{p}_f,\vec{p}_f^{\prime})
\hat f^K(\vec{p}_f^{\prime},\vec{R};\epsilon,t)\big>
\enspace ,
\end{equation}
\begin{equation}\label{nuequation}
\hat\nu^{R,A}_{mf}(\vec{p}_f,\vec{R};t)= N_f
\int_{-\epsilon_c}^{+\epsilon_c}{d\epsilon\over 4\pi i}
\big<
A(\vec{p}_f,\vec{p}_f^{\prime})
\hat{\mbox{g}}^K(\vec{p}_f^{\prime},\vec{R};\epsilon,t)\big>
\enspace ,
\end{equation}
\begin{equation}\label{Kequations}
\hat\Delta^{K}_{mf}(\vec{p}_f,\vec{R};t)=0\enspace\,,
\qquad
\hat\nu^{K}_{mf}(\vec{p}_f,\vec{R};t)=0\enspace .
\end{equation}
The impurity self-energy, 
\begin{equation}\label{born}
\check\sigma_{i} (\vec{p}_f ,\vec{R};\epsilon,t)=
n_i \; \check{t}(\vec{p}_f,\vec{p}_f,\vec{R};\epsilon, t)
\,,
\end{equation}
is specified by the impurity concentration, $n_i$, and impurity 
scattering $t$-matrix, which is obtained from the the self-consistent 
solution of the $t$-matrix equations,
\ber\label{tmatrix}
\check{t} (\vec{p}_f,\vec{p}''_f,\vec{R};\epsilon, t) &=&
\check{u}(\vec{p}_f,\vec{p}''_f) 
\nonumber
\\
&+& N_f \big< \check{u}(\vec{p}_f,\vec{p}'_f) 
\otimes \check{g}(\vec{p}'_f,\vec{R};\epsilon, t)
\otimes \check {t} (\vec{p}'_f,\vec{p}''_f,\vec{R};\epsilon, t) \big>
\enspace .
\eer
The Nambu matrix $\hat{f}^K$ is the off-diagonal part of $\hat g^K$, 
while $\hat{\mbox{g}}^K$ is the diagonal part in particle-hole space.
The Fermi surface average is defined by
\begin{equation}\label{average}
\big<\ldots\big>\, =\
{1\over N_f}\int{d^2\vec{p}_f^{\prime}\over (2\pi)^3\mid\!\vec{v}_f^{\prime}\!\mid}\,
\left(\ldots\right)\,,\quad
N_f=\int{d^2\vec{p}_f^{\prime}\over (2\pi)^3\mid\vec{v}_f^{\prime}\!\mid} \enspace ,
\end{equation}
where $N_f$ is the average density of states on the Fermi surface.
The other material parameters that enter the self-consistency equations are
the dimensionless pairing interaction, $N_fV(\vec{p}_f,\vec{p}_f^{\prime})$,  
the dimensionless Landau interaction, $N_fA(\vec{p}_f,\vec{p}_f^{\prime})$,
the impurity concentration, $n_i$, the impurity potential, 
$\check u(\vec{p}_f,\vec{p}'_f)$, and the Fermi surface data:
$\vec{p}_f$ (Fermi surface), $\vec{v}_f(\vec{p}_f)$ (Fermi velocity).
We eliminate both the magnitude of the pairing 
interaction and the cut-off, $\epsilon_c$,
in favor of the transition temperature, $T_c$, using
the linearized, equilibrium form of the mean-field gap 
equation (Eq. (\ref{gapequation})).

The quasiclassical equations are supplemented by constitutive 
equations for the charge density, the current density and the induced
electromagnetic potentials. The formal result for the
non-equilibrium charge density, to linear order in $\Delta/E_f$,
is given in terms of the Keldysh propagator by
\begin{equation}\label{density}
\hspace*{-2mm}
\rho^{(1)}(\vec{R};t)=
eN_f\int_{-\epsilon_c}^{+\epsilon_c}{d\epsilon\over 4\pi i}
\big<Z(\vec{p}'_f)\,\mbox{Tr}\,
\left[\hat g^K(\vec{p}_f^{\prime},\vec{R};\epsilon,t)\right]\big>
-2e^2N_fZ_0\Phi(\vec{R};t)
\,,
\end{equation}
with the renormalization factors given by
\be\label{renormalization}
Z(\vec{p}_f)=1-\big\langle A(\vec{p}'_f,\vec{p}_f )\big\rangle
\,,\quad
Z_0= \big\langle Z(\vec{p}_f) \big\rangle
\,.
\ee
The high-energy renormalization factor is related to an average 
of the scattering amplitude on the Fermi surface by a Ward identity 
that follows from the conservation law for charge \cite{ser83}.
The charge current induced by $\vec{A}(\vec{R};t)$, calculated to leading 
order in $\Delta/E_f $, is also obtained from the Keldysh 
propagator,
\begin{equation}\label{current1}
\vec{j}^{(1)}(\vec{R};t)= eN_f
\int {d\epsilon\over
4\pi i}\mbox{Tr}\big<\vec{v}_f(\vec{p}_f^{\prime})\hat\tau_3
\hat g^K(\vec{p}_f^{\prime},\vec{R};\epsilon,t) \big>
\,.
\end{equation}
There is no additional high-energy renormalization of the 
coupling to the vector potential because the quasiparticle 
Fermi velocity already includes the high-energy renormalization 
of the charge-current coupling in Eq. \ref{current1}.
Furthermore, the self-consistent solution of the quasiclassical 
equations for $\hat{g}^K$ ensures the continuity equation for 
charge conservation,
\begin{equation}
\label{conserv1}
\partial_t\;\rho^{(1)}(\vec{R};t)+\grad \cdot \vec{j}^{(1)}(\vec{R};t) = 0
\,,
\end{equation}
is satisfied.

An estimate of the contribution to the charge density 
from the integral in Eq. (\ref{density})
leads to the condition of ``local charge neutrality'' \cite{gor76,art79}.
A charge density given by the elementary charge times the number of states 
within an energy interval $\Delta$ around the 
Fermi surface implies $\rho^{(1)}\sim 2eN_f\Delta $.
Such a charge density cannot be maintained within 
a coherence volume because of the cost in Coulomb energy.
The Coulomb energy is suppressed by requiring the leading
order contribution to the charge density vanish: i.e. 
$\rho^{(1)}(\vec{R};t)=0$. Thus, the spatially varying
renormalized electro-chemical potential, $Z_0\Phi$, is determined by
\be\label{localneutral}
2eZ_0\Phi(\vec{R};t)=
\int_{-\epsilon_c}^{+\epsilon_c}{d\epsilon\over
4\pi i}\mbox{Tr}\big<
Z(\vec{p}'_f)
\hat g^K(\vec{p}_f^{\prime},\vec{R};\epsilon,t)\big>
\,.
\ee
The continuity equation implies $\grad \cdot\vec{j}^{(1)}(\vec{R};t)=0$.
We discuss violations of the charge neutrality condition (\ref{localneutral}),
which are higher order in $\Delta/E_f$, in Sec. \ref{chargeresp}. 
Finally, Ampere's equation, with the current given by 
Eq. (\ref{current1}), determines the
vector potential in the quasiclassical approximation,
\begin{equation}\label{vecpot}
\rot \rot \vec{A}(\vec{R};t)=
\frac{8\pi eN_f}{c}\int{d\epsilon\over 4\pi i}\mbox{Tr}\big<
\vec{v}_f(\vec{p}'_f) \hat\tau_3
\hat g^K(\vec{p}_f^{\prime},\vec{R};\epsilon,t)\big>
\enspace .
\end{equation}

Equations (\ref{transpqcl})-(\ref{tmatrix}) and 
(\ref{localneutral})-(\ref{vecpot}) constitute
a complete set of equations for calculating the 
electromagnetic response of vortices in the quasiclassical 
limit. For high-$\kappa$ superconductors we can simplify 
the self-consistency calculations to some degree.
Since quasiparticles couple to the vector potential via
$\frac{e}{c} \vec{v}_f \cdot \vec{A}$, Eq. (\ref{vecpot}) shows that
this quantity is of order $8\pi e^2 N_fv_f^2/c^2=1/\lambda^2$,
where $\lambda$ is the magnetic penetration depth. 
Thus, for $\kappa=\lambda/\xi_0\gg 1$, as in the layered cuprates, 
the feedback effect of the current density on the vector potential 
is small by factor $1/\kappa^2$.

\subsection{Linear Response}\label{Sect_linresp}

For sufficiently weak fields we can calculate the electromagnetic response
to linear order in the external field. The propagator and the
self-energies are separated into unperturbed equilibrium parts 
and terms that are first-order 
in the perturbation,
\begin{equation}\label{linresp}
\check{g}=\check{g}_0+\delta\check{g}\,, \,\,
\check{\Delta}_{mf}=\check{\Delta}_{0} +\delta\check{\Delta}_{mf}\,, \,\,
\check{\sigma}_{i}=\check{\sigma}_{0}+\delta\check{\sigma}_{i}
\enspace ,
\end{equation}
and similarly for the electromagnetic potentials,
$\vec{A}=\vec{A}_0+\delta\vec{A}$, $\Phi=\delta\Phi$.
The equilibrium propagators obey the matrix transport equation,
\begin{equation}\label{transpqcl0}
\bigl[(\epsilon+{e\over c}\vec{v}_f\cdot\vec{A}_0)
\check\tau_3-\check\Delta_{0}-\check\sigma_{0}\,,\check
g_0\bigr] +i\vec{v}_f\cdot\grad \check g_0\\ = \ 0
\,.
\end{equation}
These equations are supplemented by the 
self-consistency equations for the mean fields,
Eqs. (\ref{gapequation})-(\ref{nuequation}), 
the impurity self energy, Eqs. (\ref{born})-(\ref{tmatrix}),
the local charge-neutrality condition for the scalar potential, 
Eq. (\ref{localneutral}), Amp\`ere's equation for the vector 
potential, Eq. (\ref{vecpot}),
the equilibrium normalization conditions,
\begin{equation}\label{normalize0}
\check{g}_0^2=-\pi^2\check{1}
\,,
\end{equation}
and the equilibrium relation between the Keldysh function 
and equilibrium spectral density,
\begin{equation}\label{KMS}
\hat{g}^K_0=\tanh\left(\frac{\epsilon}{2T}\right)
            \Big[\hat{g}^R_0-\hat{g}^A_0\Big]
\,.
\end{equation}

The first-order correction to the matrix propagator 
obeys the linearized transport equation,
\begin{equation}\label{transpqcl1}
\bigl[(\epsilon+{e\over c}\vec{v}_f\cdot\vec{A}_0)
\check\tau_3-\check\Delta_{0}-\check\sigma_{0}\,,\delta\check
g\bigr]_{\otimes} +i\vec{v}_f\cdot\grad \delta\check g\\ = \
\bigl[
\delta\check{\Delta}_{mf}+\delta\check{\sigma}_{i}+\delta\check{v}\,,\
\check{g}_0]_{\otimes}\enspace ,
\end{equation}
with source terms on the right-hand side from both 
the external field ($\delta\check{v}$) and the internal 
fields ($\delta\check{\Delta}_{mf}$, $\delta\check{\sigma}_i$). In addition,
the first-order propagator satisfies the ``orthogonality condition'',
\begin{equation}\label{normalize1}
\check{g}_0\otimes\delta\check{g}
+\delta\check{g}\otimes\check{g}_0=0\enspace .
\end{equation}
obtained from linearizing the full normalization condition.\footnote{
Note that the convolution product between an equilibrium and a 
nonequilibrium quantity simplifies after Fourier transforming
$t\rightarrow\omega$:
$\check A_0 \otimes \delta \check B (\epsilon ,\omega )
= \check A_0 (\epsilon + \omega/2 )\check B (\epsilon ,\omega )$, 
$\check B (\epsilon ,\omega ) \otimes \check A_0 
= \check B (\epsilon ,\omega ) \check A_0 (\epsilon - \omega/2 )$.}
The system of linear equations are supplemented
by the equilibrium and first-order self-consistency conditions for the 
order parameter,
\begin{equation}\label{lgapequation0}
\hat\Delta^{R,A}_{0}(\vec{p}_f,\vec{R})=
N_f\int_{-\epsilon_c}^{+\epsilon_c}{d\epsilon\over 4\pi i}
\big< V(\vec{p}_f,\vec{p}'_f)
\hat f^K_0(\vec{p}_f^{\prime},\vec{R};\epsilon)\big>
\,,
\end{equation}
\begin{equation}\label{lgapequation1}
\delta\hat\Delta^{R,A}_{mf}(\vec{p}_f,\vec{R};t)=
N_f\int_{-\epsilon_c}^{+\epsilon_c}{d\epsilon\over 4\pi i}
\big< V(\vec{p}_f,\vec{p}'_f)
\delta\hat f^K(\vec{p}_f^{\prime},\vec{R};\epsilon,t)\big>,
\end{equation}
and the impurity self-energy,
\ber\label{respborn0}
\hspace*{-5mm}
\check\sigma_{0} (\vec{p}_f ,\vec{R};\epsilon) &=&
n_i\;\check{t}_0(\vec{p}_f,\vec{p}_f,\vec{R};\epsilon )\,,
\\
\hspace*{-5mm}
\label{tmatrix0}
\check{t}_0 (\vec{p}_f,\vec{p}''_f,\vec{R};\epsilon )&=&
\check {u} (\vec{p}_f,\vec{p}''_f)+
N_f \big< \check {u} (\vec{p}_f,\vec{p}'_f)
\check{g}_0(\vec{p}'_f,\vec{R};\epsilon )
\check {t}_0 (\vec{p}'_f,\vec{p}''_f,\vec{R};\epsilon ) \big>\,,
\\
\hspace*{-5mm}
\label{tmatrix1}
\delta \check\sigma_{i} (\vec{p}_f ,\vec{R};\epsilon,t)&=&
n_i N_f \big< \check {t}_0 (\vec{p}_f,\vec{p}'_f,\vec{R};\epsilon )
\otimes \delta \check{g}(\vec{p}'_f,\vec{R};\epsilon, t)
 \otimes \check {t}_0 (\vec{p}'_f,\vec{p}_f,\vec{R};\epsilon ) \big>
\,.
\eer

In general the diagonal mean fields also contribute to the response. 
However, we do not expect Landau interactions to lead to qualitatively 
new phenomena for the vortex dynamics, so we have neglected these 
interactions in the following analysis
and set $A(\vec{p}_f,\vec{p}_f')=0$ (i.e. $\check \nu_{mf}=0$).
As a result the local charge neutrality condition for the 
electro-chemical potential becomes,
\be\label{resplocalneutral}
2e\delta\Phi(\vec{R};t)=\int_{-\epsilon_c}^{+\epsilon_c}{d\epsilon\over
4\pi i}\mbox{Tr}\big< \delta \hat g^K(\vec{p}_f,\vec{R};\epsilon,t)\big>
\,.
\ee

In what follows we work in a gauge in which the induced 
electric field, $\vec{E}^{\mbox{\tiny ind}}(\vec{R};t)$,
is obtained from $\delta\Phi(\vec{R};t)$ and the uniform 
external electric field, $\vec{E}_{\omega}^{\mbox{\tiny ext}}(t)$,
is determined by the vector potential $\delta\vec{A}_\omega (t)$.
For $\lambda/\xi_0\gg 1$ we can safely neglect corrections 
to the vector potential due to the induced current. 
Thus, in the Nambu-Keldysh matrix notation the electromagnetic 
coupling to the quasiparticles is given by
\begin{equation}\label{perturbation}
\delta\check{v} =
-{e\over c}\vec{v}_f\cdot\delta\vec{A}_\omega (t)\check{\tau}_3
+e\delta\Phi(\vec{R};t)\check{1}
\enspace .
\end{equation}
The validity of linear response theory requires the external
perturbation $\delta \check{v}$ be sufficiently small and that
the induced vortex motion responds to the external field at
the frequency set by the external field. At very low frequencies 
frictional damping of the vortex motion, arising from the finite mean 
free path of quasiparticles scattering from impurities, gives rise to a 
nonlinear regime in the dynamical response of a vortex.
This regime is discussed extensively in the literature \cite{lar77},
and is not subject of our study. However, for sufficiently small 
field strengths the vortex motion is nonstationary over any time interval, 
although it may be regarded as quasi-stationary at low enough frequencies.
The nonstationary motion of the vortex can be described by
linear response theory if $\delta\check{v}\ll 1/\tau$ for
$\omega\lsim 1/\tau$, and $\delta \check{v}\ll \omega$ for
$\omega\gsim 1/\tau$. Note that the frequency of the perturbation, 
$\omega$, is not required to be small compared to the gap
frequency; it is only restricted to be small compared to 
atomic scale frequencies, e.g $\omega \ll E_f/\hbar$.

Self-consistent solutions of Eqs. (\ref{lgapequation1}), (\ref{tmatrix1}) 
and (\ref{resplocalneutral}) for the self-energies and scalar potential 
are  fundamental to obtaining a physically sensible solution for the 
electromagnetic response. The dynamical self-energy corrections
are equivalent to ``vertex corrections'' in the Kubo formulation of linear 
response theory. They are particularly important in the context of 
nonequilibrium phenomena in inhomogeneous 
superconductors.\footnote{Vertex corrections usually vanish in
homogeneous superconductors because of translational and rotational symmetries.
Inhomogeneous states break these symmetries and 
typically generate non-vanishing vertex corrections.}
In our case these corrections are of vital importance; 
the self-consistency conditions enforce charge conservation.
In particular, Eqs. (\ref{respborn0})-(\ref{tmatrix1}) imply charge
conservation in scattering processes,  whereas (\ref{lgapequation0}) 
and  (\ref{lgapequation1}) imply charge conservation in particle-hole 
conversion processes; any charge which is lost (gained)
in a particle-hole conversion process is compensated by a corresponding
gain (loss) of condensate charge. It is the coupled quasiparticle and 
condensate dynamics which conserves charge in superconductors. 
Neglecting the dynamics of either component, or using a 
non-conserving approximation for the coupling leads to unphysical results. 

Self-consistent calculations for the equilibrium order parameter, 
impurity self-energy and local excitation spectrum (spectral density)
are necessary inputs to the linearized transport equations for the 
dynamical response of a vortex. The equilibrium spectral function also provides 
key information for the interpretation of the dynamical response. 
Because of particle-hole coherence
the spectral density is sensitive to the phase winding 
and symmetry of the order parameter, as well as material properties 
such as the transport mean-free path and impurity cross-section.
In the following section we present results for the low-energy 
excitation spectra of singly- and doubly-quantized vortices in 
layered superconductors with $s$-wave and $d$-wave pairing  symmetry.

\section{Electronic Structure of Vortices}\label{Sec:Electronic_Structure}

The local density of states for excitations with Fermi momentum 
$\vec{p}_f$ is obtained from the retarded and 
advanced quasiclassical propagators,
\be
N(\vec{p}_f,\vec{R};\epsilon )= N_f \frac{1}{4\pi i} \mbox{Tr}
\left[\hat \tau_3\hat g^A_0(\vec{p}_f,\vec{R};\epsilon)-
\hat\tau_3\hat g^R_0(\vec{p}_f,\vec{R};\epsilon) \right]
\,.
\ee
This function measures the local density of the quasiparticle
states with energy $\epsilon$ at the point $\vec{p}_f$ on the
Fermi surface. The local density of states (LDOS) is obtained 
by averaging this quantity over all momentum directions of 
the quasiparticles,
\be
N(\vec{R};\epsilon)=\left<N(\vec{p}'_f,\vec{R};\epsilon )\right>
\,.
\ee
The product of the angle-resolved density of states and the
Fermi velocity, $\vec{v}_f$, determines the current density
carried by these states; $\vec{v}_f$ also defines
the direction of a quasiclassical trajectory passing
through the space point $\vec{R}$. We introduce the 
angle-resolved {\sl spectral current density} \cite{rai96},
\be
\vec{j}(\vec{p}_f,\vec{R};\epsilon )=
2e\vec{v}_f(\vec{p}_f)\,N(\vp_f,\vR;\epsilon)
\,,
\ee
which measures the current density carried by
quasiparticle states with energy $\epsilon$
moving along the trajectory defined by $\vec{v}_f$.
The local spectral current density is then defined as
\be
\vec{j}(\vec{R};\epsilon)=
\left<\vec{j}(\vec{p}'_f,\vec{R};\epsilon )\right>
\,,
\ee
and the total current density is obtained by summing over the
occupied states for each trajectory,
\be
\vec{j}(\vR)=\int d\epsilon\,f(\epsilon)\,\vec{j}(\vR;\epsilon)
\,,
\ee
where $f(\epsilon)=1/(1+e^{\beta\epsilon})$.
Self-consistent calculations of the equilibrium 
structure and spectral properties  of vortices are 
relatively straight-forward computations.
Below we present results for $s$-wave and $d$-wave 
pairing symmetry with impurity scattering included.

The calculations reported are carried for a circular Fermi surface, 
with an isotropic Fermi momentum $\vec{p}_f$ and Fermi velocity 
$\vec{v}_f$. The elastic scattering rate is chosen to represent 
the intermediate-clean regime, $\Delta^2/E_f < \hbar/\tau \ll \Delta$.
The pairing potential can be represented as a sum over invariant products 
of basis functions $\{\eta_{\Gamma,i}(\vec{p}_f) | i=1\dots
d_{\Gamma}\}$ for the irreducible representations of the crystal point 
group labeled by $\Gamma$,
\be
N_f V(\vec{p}_f,\vec{p}'_f)= \sum_{\Gamma,i}
v_\Gamma^{\,}\eta_{\Gamma i}^\ast(\vec{p}_f)\eta_{\Gamma i}(\vec{p}'_f)
\,.
\ee
The pairing interaction, $v_\Gamma$, and the cutoff, $\epsilon_c$, are
eliminated in favor of the instability temperature for 
pairing in symmetry channel $\Gamma$. We limit the discussion here to 
even-parity, one-dimensional representations, which for tetragonal 
symmetry includes the `$s$-wave' (identity) representation, A$_{1g}$, 
two `$d$-wave' representations, B$_{1g}$ and B$_{2g}$, and
a  `$g$-wave' representation, A$_{2g}$. The corresponding
basis functions we use are listed in Table \ref{table:basis_functions}.

\begin{table}
\begin{center}
\begin{minipage}{0.75\hsize}
\begin{center}
\noindent\caption{Symmetry classes and model basis functions for the 1D 
even-parity representations of $D_{4h}$. The angle $\phi$ is the angular
position of $\vp_f$ on the Fermi surface with respect to the 
crystallographic $a$-axis (=$x$-axis).}
\renewcommand{\arraystretch}{1.4}
\setlength\tabcolsep{5pt}
\begin{tabular}{|c|c|c|}
\hline\noalign{\smallskip}
Pairing Symmetry & Representation [$\Gamma$] &Basis Function [$\eta_{\Gamma}$] \\
\hline
$s$-wave & $A_{1g}$ &           1            \\
\hline
$d$-wave & $B_{1g}$ & $\sqrt{2}\cos(2\phi)$  \\
\hline
$d'$-wave & $B_{2g}$ & $\sqrt{2}\sin(2\phi)$ \\
\hline
$g$-wave & $A_{2g}$ & $\sqrt{2}\sin(4\phi)$  \\
\hline
\end{tabular}
\label{table:basis_functions}
\end{center}
\end{minipage}
\end{center}
\end{table}

The results for the order parameter, impurity self energy 
and spectral properties of vortices that follow are calculated self 
consistently in the $t$-matrix approximation for point impurities 
(pure $s$-wave scattering), i.e. $\check{u}(\vec{p}_f,\vec{p}'_f)=u_0\check{1}$. 
The quasiparticle scattering rate, $1/2\tau$, and normalized impurity 
cross section, $\bar{\sigma}$, are then given by,
\begin{equation}
\frac{1}{2\tau} = \frac{n_i}{\pi N_f}\,\bar{\sigma}
\,,\qquad
\bar{\sigma} = \frac{(\pi N_f u_0)^2}{1+(\pi N_f u_0)^2}
\,.
\end{equation}
The Born limit corresponds to $\bar{\sigma}\ll 1$,
while unitary scattering corresponds to
$|u_0|\rightarrow\infty$ or $\bar{\sigma} \rightarrow 1$.
In the calculations that follow 
the temperature is set at $T=0.3T_c$, and the mean free path is 
is chosen to represent the intermediate-clean regime;
$\ell=10\,\xi_0$.

\subsection{Singly quantized vortices for S-wave Pairing}

For isotropic $s$-wave pairing the equilibrium order parameter for an 
isolated vortex with winding number $p$ has the form, 
\be
\Delta(\vec{p}_f,\vec{R}) = \vert\Delta(\vec{R})\vert e^{ip\varphi}
\,,
\ee
where the amplitude $\vert\Delta(\vec{R})\vert$ is isotropic 
and $\varphi$ is the azimuthal angle of $\vec{R}$ in the plane.

\begin{figure}
\begin{center}
\leavevmode
\includegraphics[width=.49\hsize]{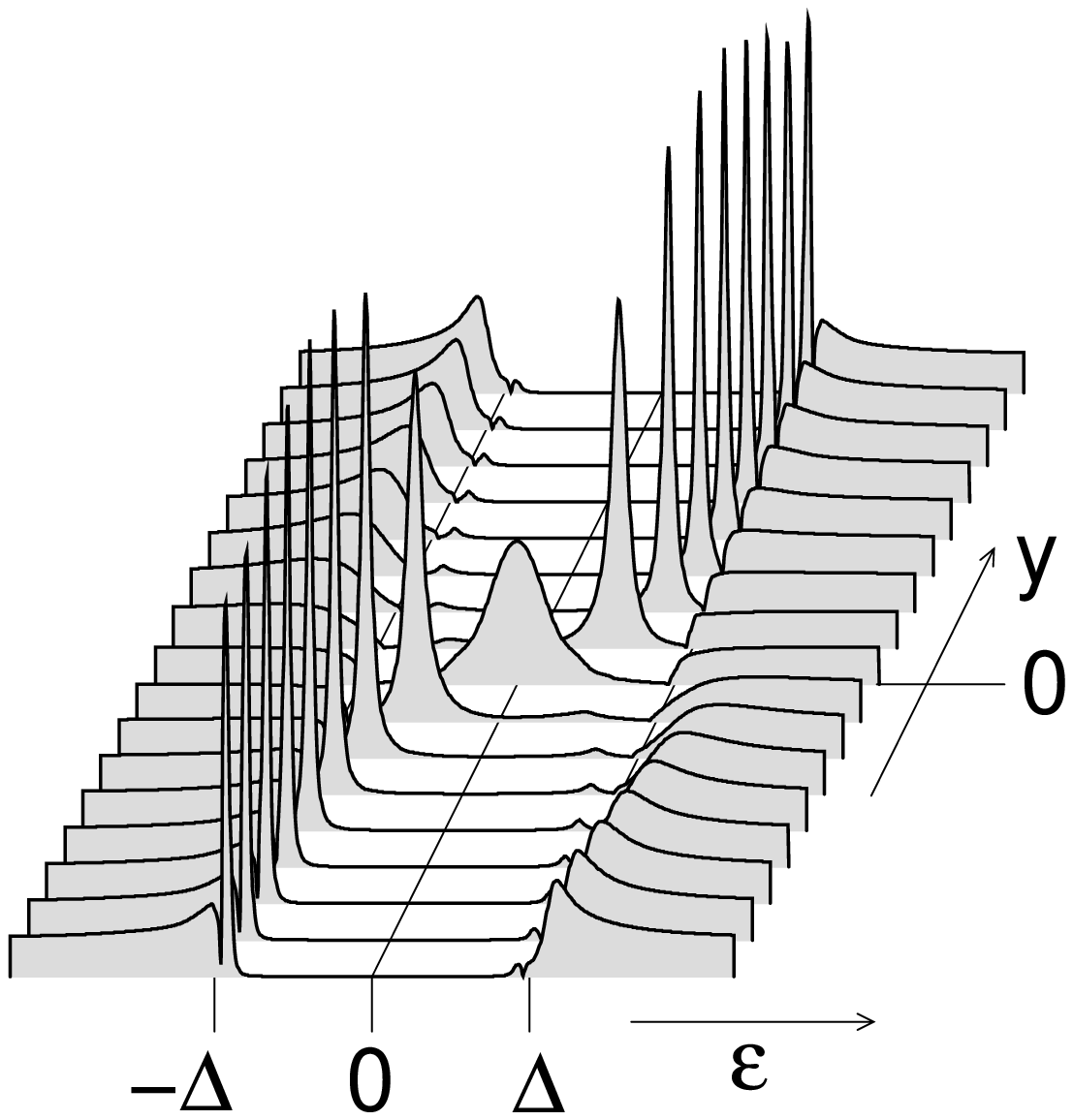}
\includegraphics[width=.49\hsize]{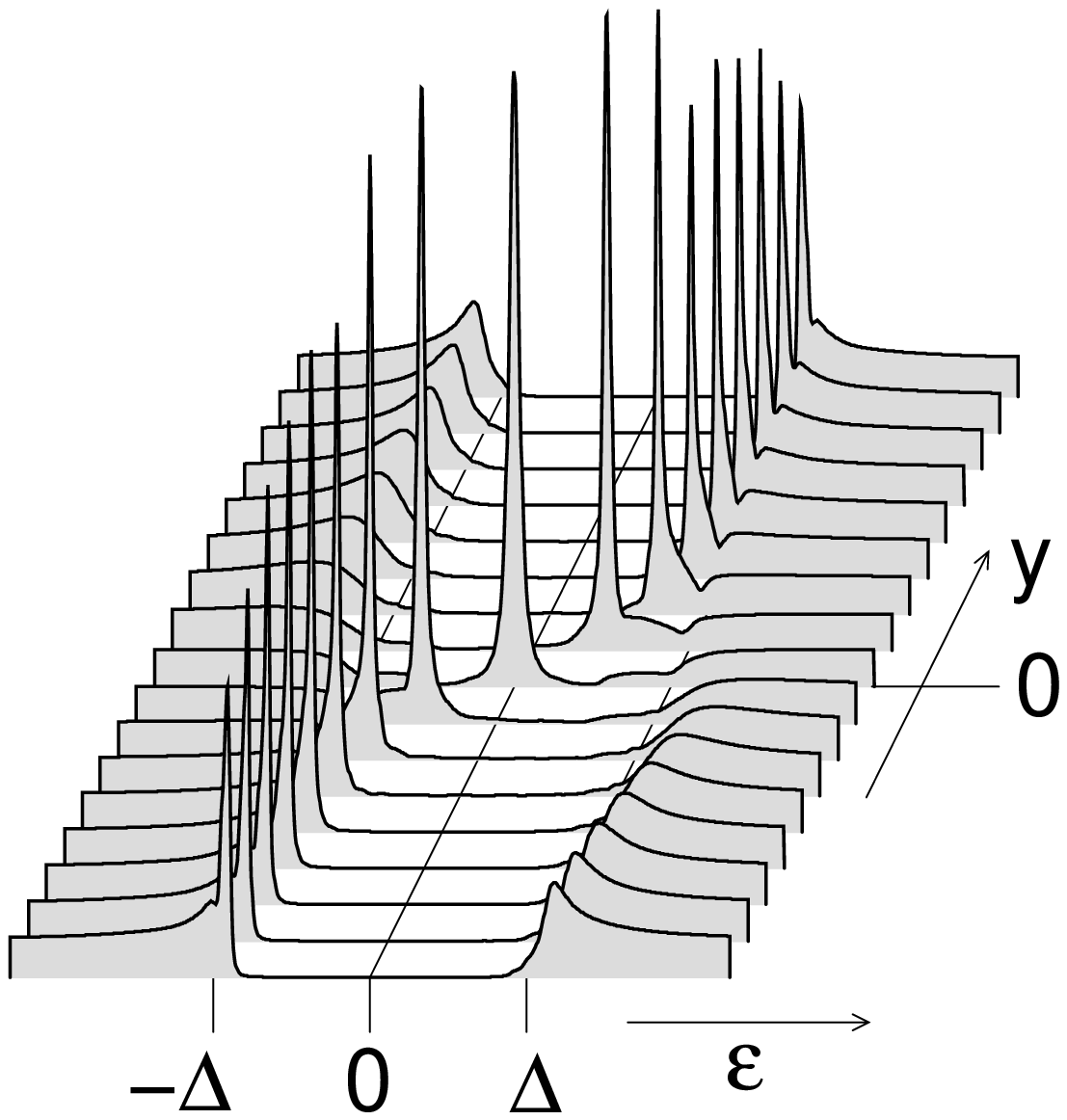}
\end{center}
\caption{ \label{Spec1a}
Angle-resolved density of states for an $s$-wave vortex 
for quasiparticles propagating along trajectories parallel 
to the $x$-direction, at points along the $y$-axis spaced by 
$1.38\,\xi_0$. The left panel is for Born scattering, and 
the right panel is for unitary scattering with the same mean 
free path of $\ell = 10\xi_0$. The temperature is $T=0.3T_c$.
}
\end{figure}

The angle-resolved local density of states spectra for a
singly quantized vortex is shown in Fig. \ref{Spec1a} for 
space points $\vec{R}=(0,y)$ along the $y$-axis with a spacing of 
$\delta y = \sqrt{3}\pi/4 \xi_0 \simeq 1.36 \xi_0$,
and for trajectories parallel to the x-axis, $\vec{v}_f=v_f \vec{e}_x$.
The vortex center is at $y=0$, and
the phase winding is such that the direction of the superflow
is in $x$-direction for spectra with negative $y$ coordinate.

Far outside the core the angle-resolved density of states resembles
the BCS density of states with a gap in the spectrum roughly between
$\epsilon=\pm\Delta$, and peaks in the spectrum near the continuum edges.
Careful inspection of Fig. \ref{Spec1a} shows that the coherence peak 
for positive energy at $y=-8\delta y \simeq -11\xi_0$ is not at 
$\epsilon = \Delta$, but is shifted to higher energy by the Doppler
effect, $\Delta\epsilon = \vv_f\cdot\vp_s$, where 
$\vp_s = \tinyonehalf\hbar\grad\vartheta - \frac{e}{c}\vA$ is the 
condensate momentum \cite{rai96}.
In a homogeneous superflow field the spectrum is the Doppler-shifted 
BCS spectrum; the Doppler shift increases
with condensate momentum until pairbreaking sets in at the bulk
critical momentum, $p_c=\Delta/v_f$. However, in the vortex core 
{\sl nonlocal effects} associated with the inhomogeneous flow field
lead to a redistribution of the spectral weight near the gap edge.
The positive energy continuum 
edge is broadened considerably compared to the square-root singularity 
in the absence of the Doppler effect. The continuum starts at
$+\Delta $ even for the Doppler-shifted spectra near
the maximal current regions. In contrast, the negative energy 
continuum edge shows sharp structures due to the accumulation 
of spectral weight in 
the region between $-\Delta $ and $-\Delta + \vv_f\cdot\vp_s$.
The sharp structure corresponds to a bound state that is separated 
from the continuum edge. The density of states at the continuum edge
drops precipitously at $\epsilon = -\Delta $. 
We emphasize that nonlocal effects lead to qualitative differences in
the spectrum near the gap edges compared to the widely used approximation
of Doppler-shifted quasiparticles in a locally homogeneous superflow \cite{fra99}.
In clean superconductors nonlocal effects dominate the spectrum.

For a homogeneous superflow the current is carried mainly by the
states that are Doppler shifted the region between $-\Delta $ and $\Delta$. 
The spectral current density shows that contributions
to the current density from states outside of this region nearly cancel. 
In the case of a vortex, the bound state that splits off from the continuum 
not only robs the Doppler-shifted continuum edge of its spectral 
weight, but the bound state also carries most of the supercurrent
in the vortex core region. At distances approaching 
the vortex center the bound state is clearly resolved and disperses through
$\epsilon=0$ at zero impact parameter. As shown in the left panel of 
Fig. \ref{Spec1a} the bound state also broadens considerably for Born 
scattering as it disperses towards the Fermi level, but
remains a sharp resonance in the limit of unitary scattering, as shown
in Fig. \ref{Spec1a}.

\begin{figure}
\begin{center}
\leavevmode
\includegraphics[width=.63\hsize]{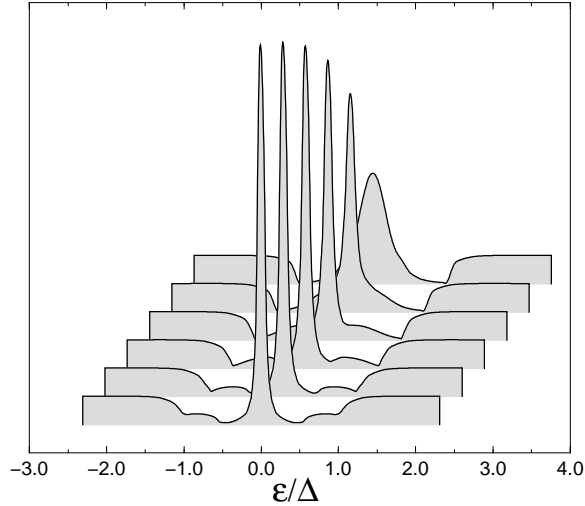}
\end{center}
\caption{ \label{unitary}
Local density of states (LDOS) in the center of an $s$-wave vortex
for different effective impurity scattering
cross sections $\bar{\sigma}$ from Born to unitary limit,
with a fixed mean free path $\ell =10\xi_0 $, and temperature $T=0.3T_c$.
From top to bottom: $\bar{\sigma} =0.0\,,0.2\,,0.4\,,0.6\,,0.8\,,1.0$. 
}
\end{figure}

The coherence peaks are completely suppressed at the vortex center,
both in the Born and unitary limits. Note the difference in the
evolution of the spectral weight for the bound and continuum states:
for positive energies the spectral weight of the coherence peak is
shifted to higher energies as one approaches the vortex core,
with a continuously diminishing intensity at the coherence peak.
In contrast, the spectral weight of the negative energy coherence peak
is transferred to the bound state which splits off from
the continuum.
In principle there could be additional, secondary bound states, which 
would split off from the continuum if the vortex core were wider.
However, a self-consistent calculation of the order parameter suppresses
the secondary bound state in the vortex core.
We return to the discussion of secondary bound states when we discuss
the spectrum of pinned vortices.
For trajectories with $y > 0$ the structure of the spectrum for positive 
and negative energies is reversed because the superflow is now counter
to the quasiparticle velocity, leading to negative Doppler shifts and 
bound states at positive energies below the continuum edge.
The small spectral features that appear at the energies corresponding 
to the {\sl negative} of the bound state energies are due to mixing of 
with trajectories of opposite velocity by backscattering from impurities.

Finally, consider the differences in the spectral features for unitary 
versus Born scattering. In Fig. \ref{unitary} we show the local density of 
states at the center of the vortex for several scattering cross sections 
ranging from the Born limit ($\bar{\sigma}\ll 1$) to
the unitary limit ($\bar{\sigma}= 1$).
In the Born limit the broadening is a maximum. For small, but 
finite cross section two bands of impurity bound states
split off from the zero-energy resonance and remove spectral weight from 
the central peak. The impurity bands evolve towards the continuum edges
as the cross-section increases and merge with them in the unitary limit.
When the impurity bound-state bands no longer overlap
the central peak the zero-energy resonance sharpens dramatically
with the width remaining constant as the unitary limit is approached.
The overlap between the impurity bands and the continuum edges in the 
unitary limit is determined by the scattering rate, and is increasing 
with increasing scattering rate, $1/\tau $. As shown in the right panel
of Fig. \ref{Spec1a}, the impurity bands are localized in the 
vortex core region; their existence depends on impurity scattering
in a region where the phase changes rapidly over length scales of
order the coherence length.

\subsection{Singly quantized vortices for D-wave Pairing}

For $d$-wave pairing symmetry the order parameter has the form,
\be
\Delta(\vp_f,\vR) = \eta_{\mbox{\tiny B$_{1g}$}}(\vp_f)\,\Delta(\vR)
\,,
\ee
where $\eta_{\mbox{\tiny B$_{1g}$}}(\vp_f)$ changes sign on the
Fermi surface at the points, $\hat{\vp}_{fx}=  \pm \hat{\vp}_{fy}$.
These nodal points lead to strong anisotropy and gapless excitations
in the quasiparticle spectrum, which feeds back to produce anisotropy
in the spatial structure of the order parameter in the core region of 
of a vortex. The spatial part of the order parameter,
\be
\Delta(\vR) = |\Delta(\vR)|\,e^{i\vartheta(\vR)}
\,,
\ee
for a vortex, at distances far from the core, approaches that of an 
isotropic vortex: $|\Delta(\vR)|\rightarrow \Delta(T)$ and 
$\vartheta\rightarrow p \varphi$. However, in the core region
the current density is comparable to the critical current density
and develops a four-fold anisotropy as a result of the backflow current 
concentrated near the nodes \cite{yip92a}. This current-induced pairbreaking 
effect is dominant for flow parallel to the nodal directions and leads to 
weak anisotropy of the order parameter in the core region.

\begin{figure}
\begin{center}
\leavevmode
\includegraphics[width=.49\hsize]{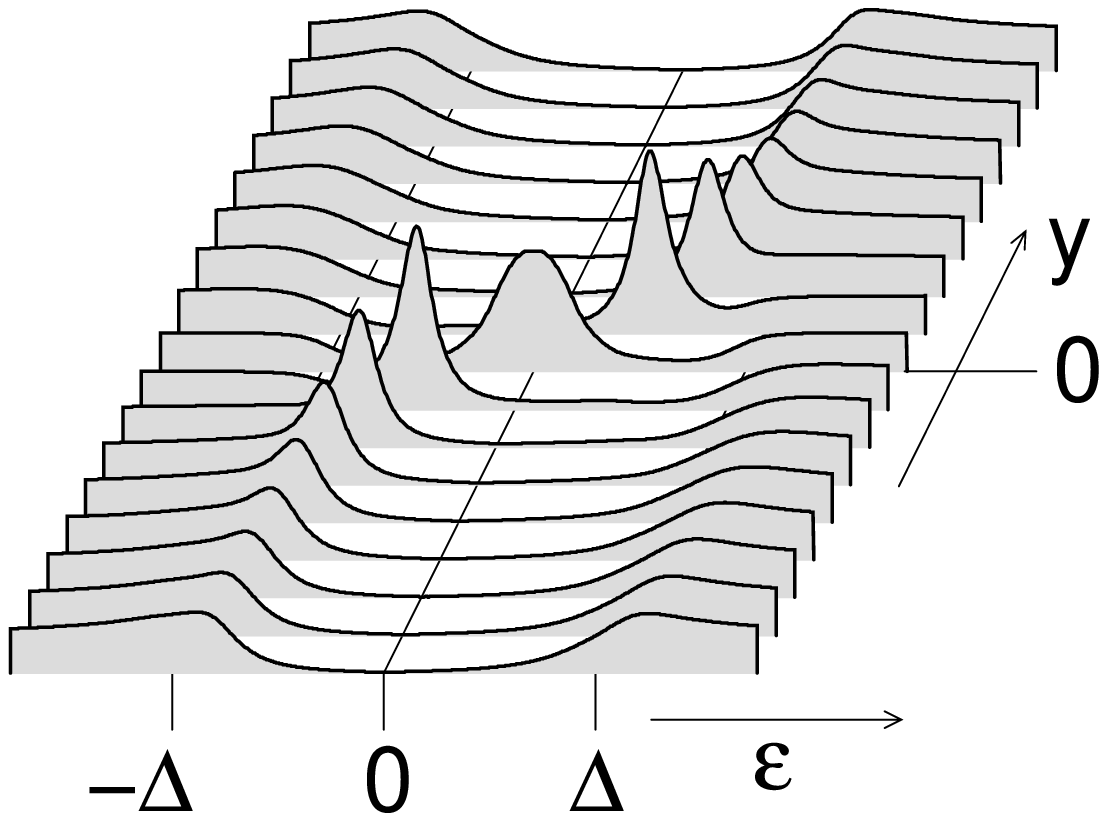}
\includegraphics[width=.49\hsize]{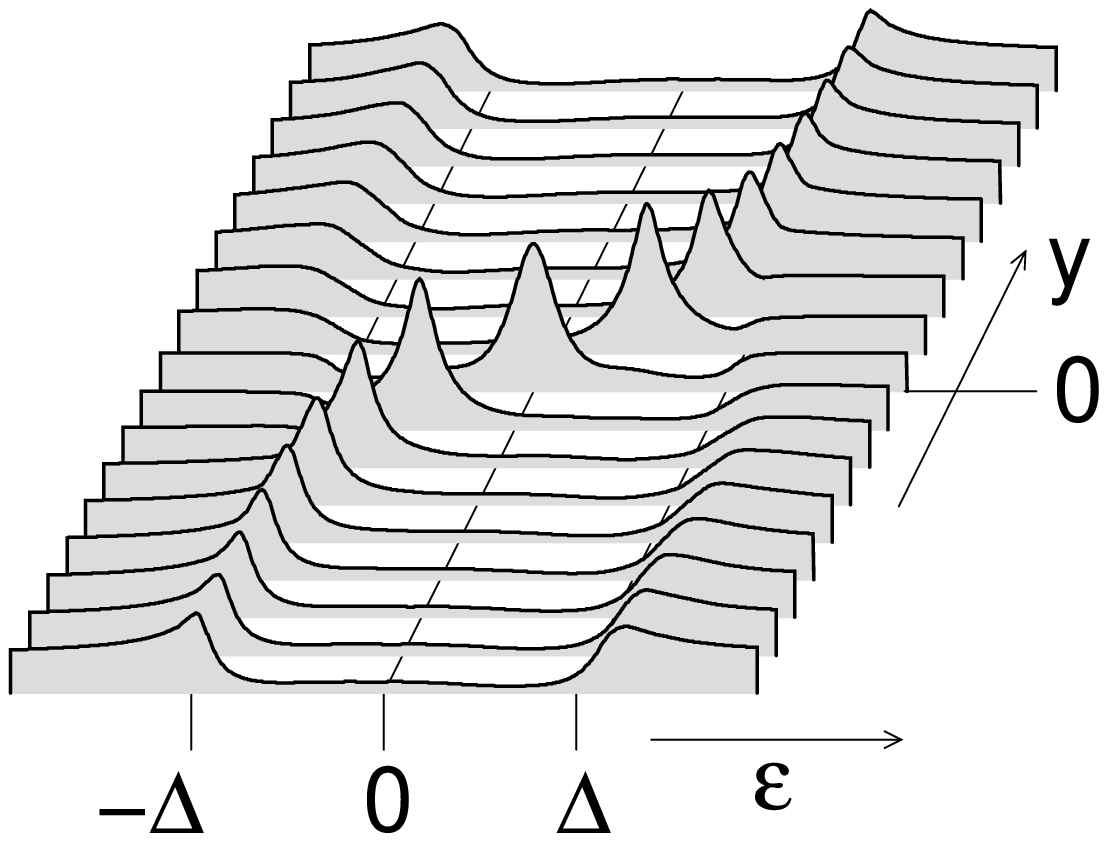}
\end{center}
\caption{ \label{Spec1d}
Angle resolved density of states for a $d$-wave vortex for a 
quasiparticle trajectory along the $x$-direction (antinodal) 
as a function of impact parameter ($y$-direction).
The impact parameter spacing is $\delta y= 1.36 \xi_0$.
The left panel corresponds to impurity scattering in the Born 
limit, and the right panel is for the unitary limit. The
temperature is $T=0.3T_c$.
}
\end{figure}

The electronic structure of the $d$-wave vortex, presented in 
Fig. \ref{Spec1d}, shows distinct differences
from that of an $s$-wave vortex.
The resonances in the vortex core are broader than the core
states of vortex in an $s$-wave superconductor.
Mixing with extended states in nodal direction
broadens the peaks near the continuum edges.
Also note the effect of impurity scattering on the spectra 
far from the vortex core region, which show a broadened
continuum rather than a sharp continuum edge.
In the right panel of Fig. \ref{Spec1d} we show the angle-resolved
density of states for a $d$-wave vortex for unitary impurity scattering.
As in the $s$-wave case there is sharpening of the zero resonance,
but note that the width of the resonance is much broader than in the case
of $s$-wave pairing because impurity scattering provides a coupling 
and mixing of the bound state resonance with the low-energy 
extended states for momenta near the nodal directions,
even for a trajectory in the antinodal direction. 

\begin{figure}
\begin{center}
\leavevmode
\includegraphics[width=.44\hsize]{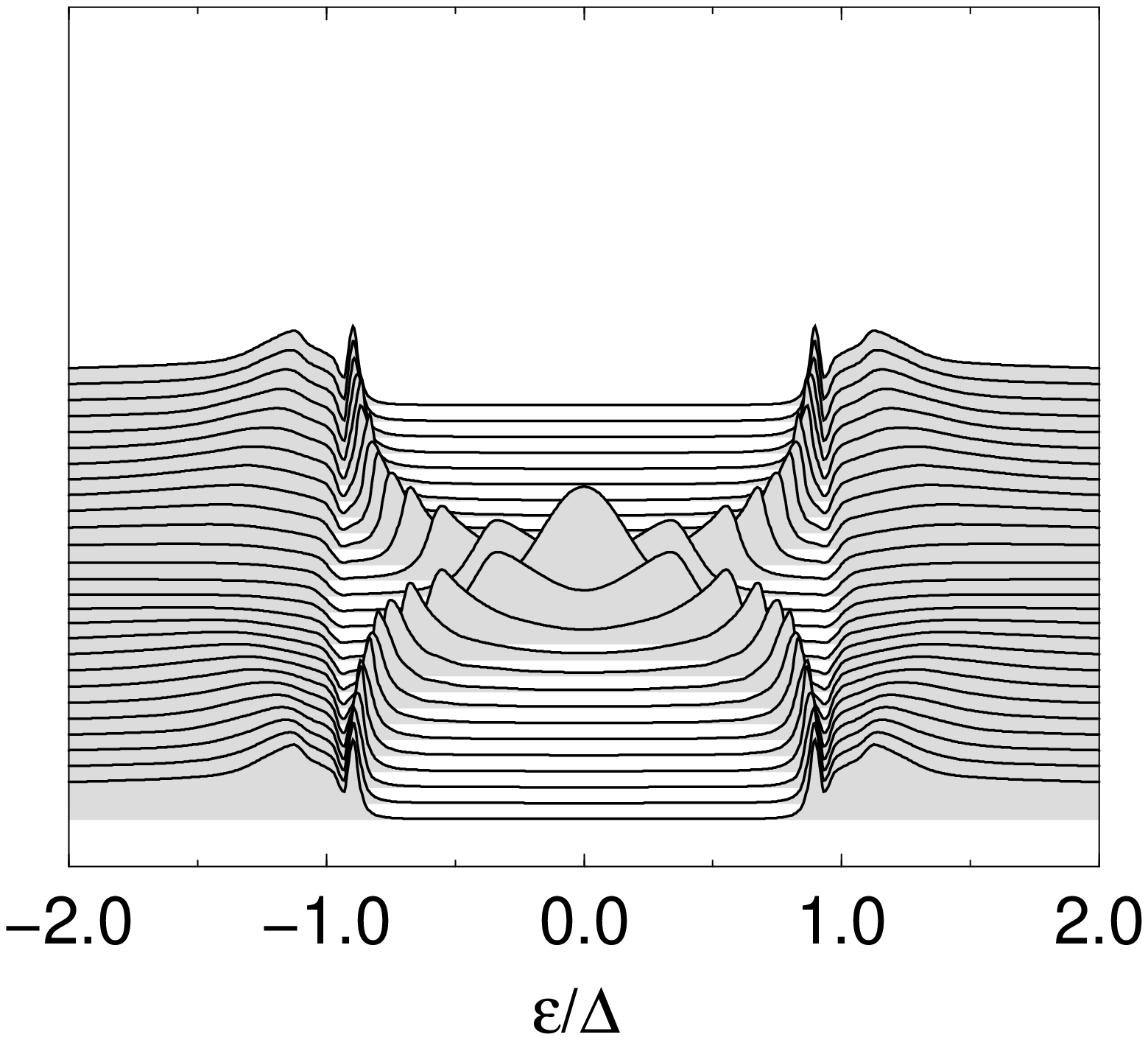}
\includegraphics[width=.44\hsize]{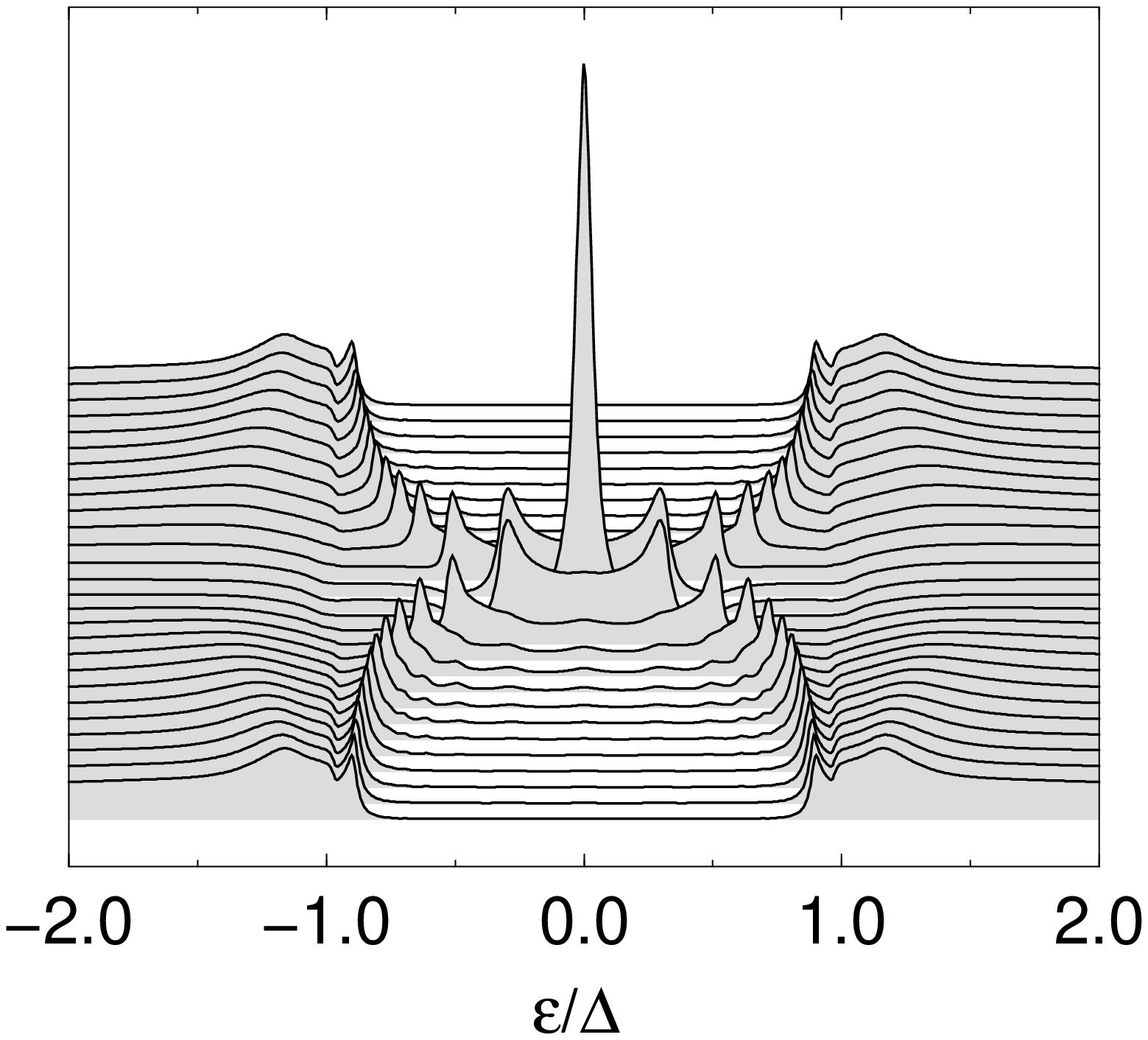}\\
\includegraphics[width=.44\hsize]{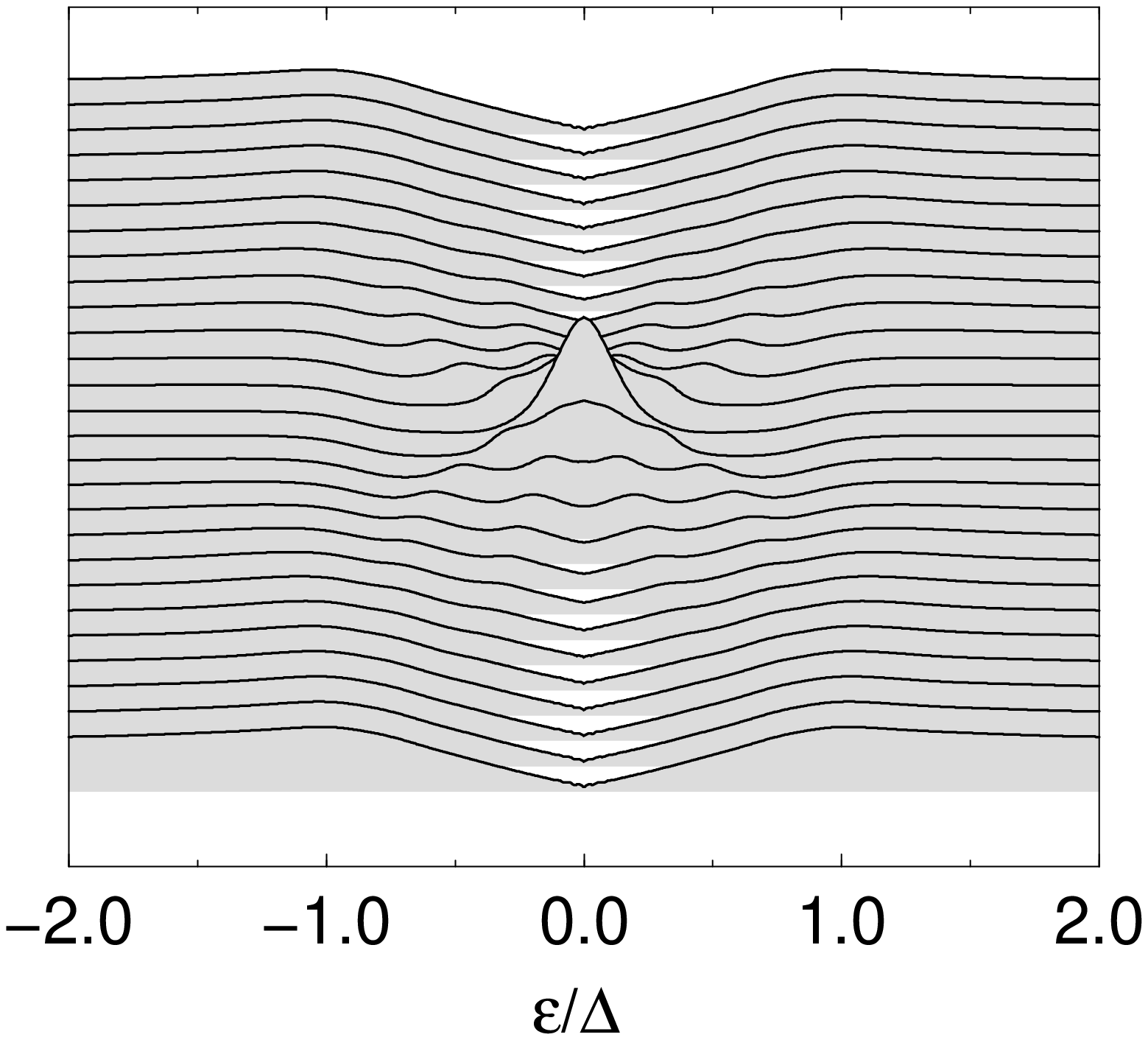}
\includegraphics[width=.44\hsize]{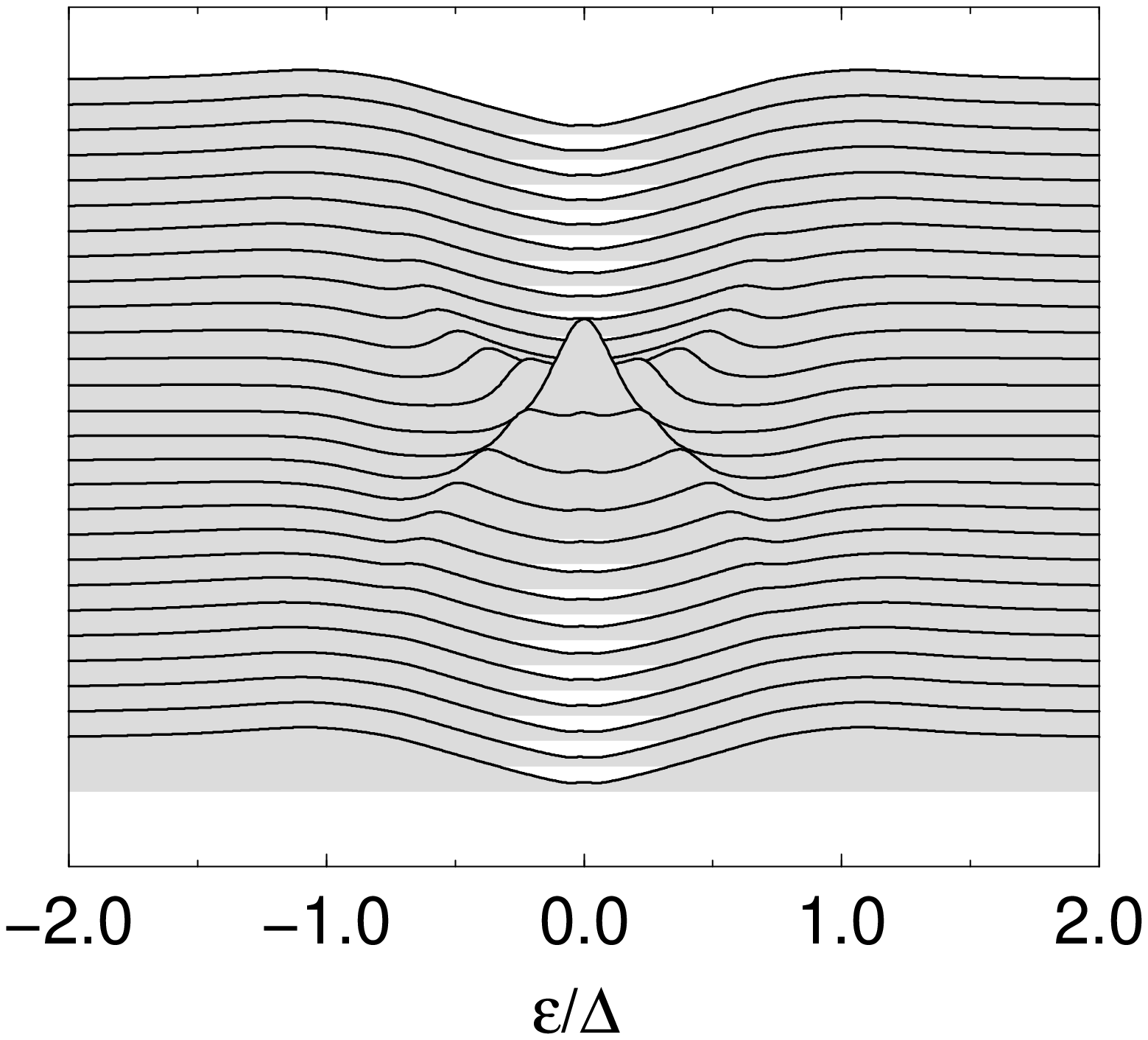}
\end{center}
\caption{ \label{Nofepss}
Local density of states for an $s$-wave vortex (top panels) and
a $d$-wave vortex (bottom panels), as a function of distance from 
the vortex center with a spacing $\delta y = 0.79 \xi_0$.
For $s$-wave pairing the Born limit is shown on the left and 
unitary scattering is on the right.
For $d$-wave pairing the LDOS is shown only for Born scattering,
but along axes parallel to a anti-nodal direction (left) and parallel
to the nodal (right). The temperature is $T=0.3T_c$.
}
\end{figure}

In Fig. \ref{Nofepss} we show the local density of states (LDOS) for
both $s$-wave (top two panels) and $d$-wave pairing (bottom two panels)
as a function of position $\vec{R}$ along an 
anti-nodal direction (bottom left) and along a nodal direction (bottom right).
The LDOS is obtained by averaging the angle-resolved density of states 
over the Fermi surface at a particular space point $\vR$. The averaging, 
together with the dispersion of the bound states and resonances as a 
function of angle leads to one-dimensional bands with characteristic 
Van Hove singularities, which are clearly visible for $s$-wave symmetry,
but considerably smeared and broadened for $d$-wave symmetry. 
The LDOS for a vortex with $s$-wave pairing symmetry is shown for 
impurity scattering in both the Born (top left) and unitary 
limits (top right). The spectra show the characteristic bound state bands,
Van Hove singularities and the dramatic reduction in the width of the 
zero bias resonance for unitary scattering. 

Calculations of the LDOS in the superclean limit for
a vortex with $d$-wave pairing are discussed by Schopohl and 
Maki \cite{sch95a} and by Ichioka {\it et al.} \cite{ich96}.
The self-consistent calculations shown in Fig. \ref{Nofepss}
include impurity scattering in the Born limit
for space points, $\vec{R}$, along two different directions; the left
panel corresponds the LDOS measured as a function of distance along
the anti-nodal direction and the right panel is the LDOS measured along
the nodal direction.
The nodes of the order parameter for $d$-wave pairing lead to 
continuum states with energies down to the Fermi level. These states
are visible in the LDOS as the smooth background extending to zero 
energy from both positive and negative energies, even for distances far 
from the core. In the vortex core region several broad peaks disperse 
as a function of distance from the vortex center. These peaks correspond 
to broadened Van Hove singularities resulting from averaging the 
vortex core resonances over the Fermi surface for at a fixed 
position $\vec{R}$. The differences in the $d$-wave spectra
for the two directions reflects the weak fourfold anisotropy 
of the LDOS around the vortex at fixed energy.

\subsection{Vortices pinned to mesoscopic metallic inclusions}

The calculations discussed above describe the average effects 
of atomic scale impurity disorder on the spectral properties 
of a vortex. A specific defect can also act as a pinning site 
for a vortex. We model such a defect as a mesoscopic size, 
normal metallic inclusion in the superconductor.
The defect is assumed to be a circular inclusion, with
a radius, $\xi_{\mbox{\tiny pin}}$, of order the coherence length of 
the superconductor. For simplicity we assume that the metallic properties 
of the inclusion, e.g. Fermi surface parameters, are the same as those of the 
normal-state of the host superconductor. The calculations presented below
neglect normal reflection processes at the interface between the inclusion and the
host metal, but include Andreev reflection. The analysis and calculations 
can be generalized for more detailed models of a pinning center.

\begin{figure}
\begin{center}
\leavevmode
\includegraphics[width=.49\hsize]{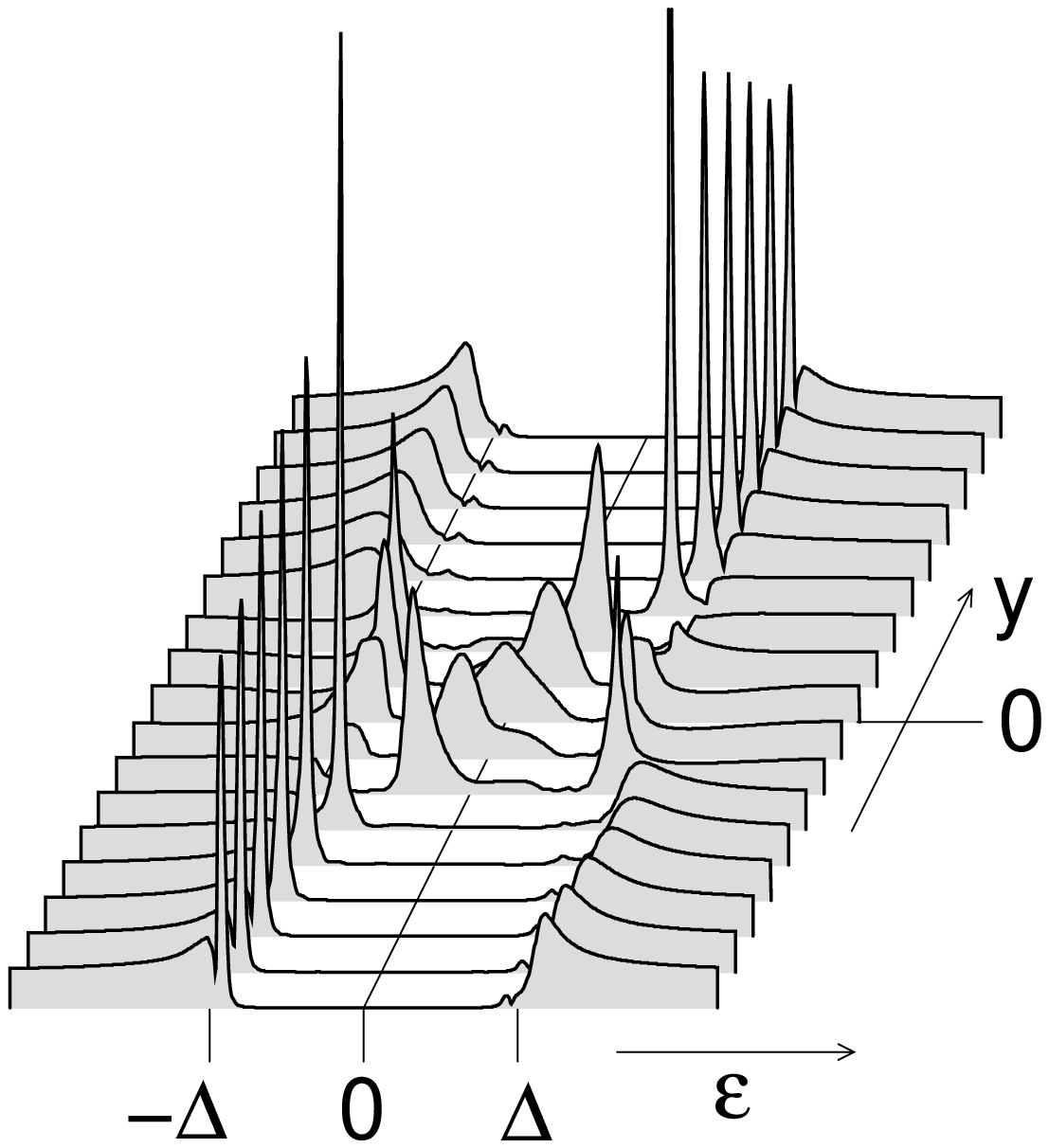}
\includegraphics[width=.49\hsize]{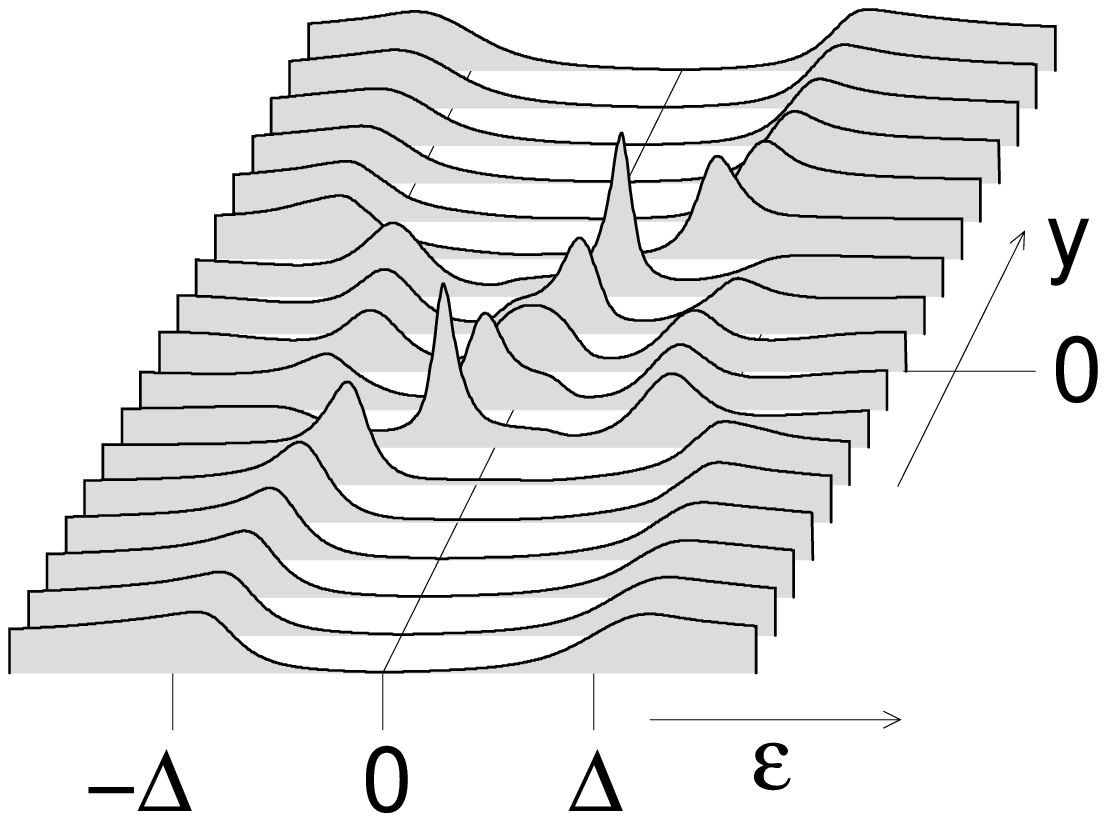}
\end{center}
\caption{ \label{ADOS_pin_s}
Angle resolved density of states for a quasiparticle trajectory
along the $x$-direction, as a function of impact parameter ($y$-axis)
with a spacing of $\delta y = 1.36 \xi_0$ for a vortex centered on 
a pinning center of radius of $\xi_{\mbox{\tiny pin}}=\pi\xi_0$.
The left panel is for $s$-wave symmetry and right panel is
for $d$-wave symmetry. Impurity scattering is included for Born scattering with 
$\ell = 10\xi_0$. The temperature is $T=0.3T_c$.
}
\end{figure}

\begin{figure}
\begin{center}
\leavevmode
\includegraphics[width=.49\hsize]{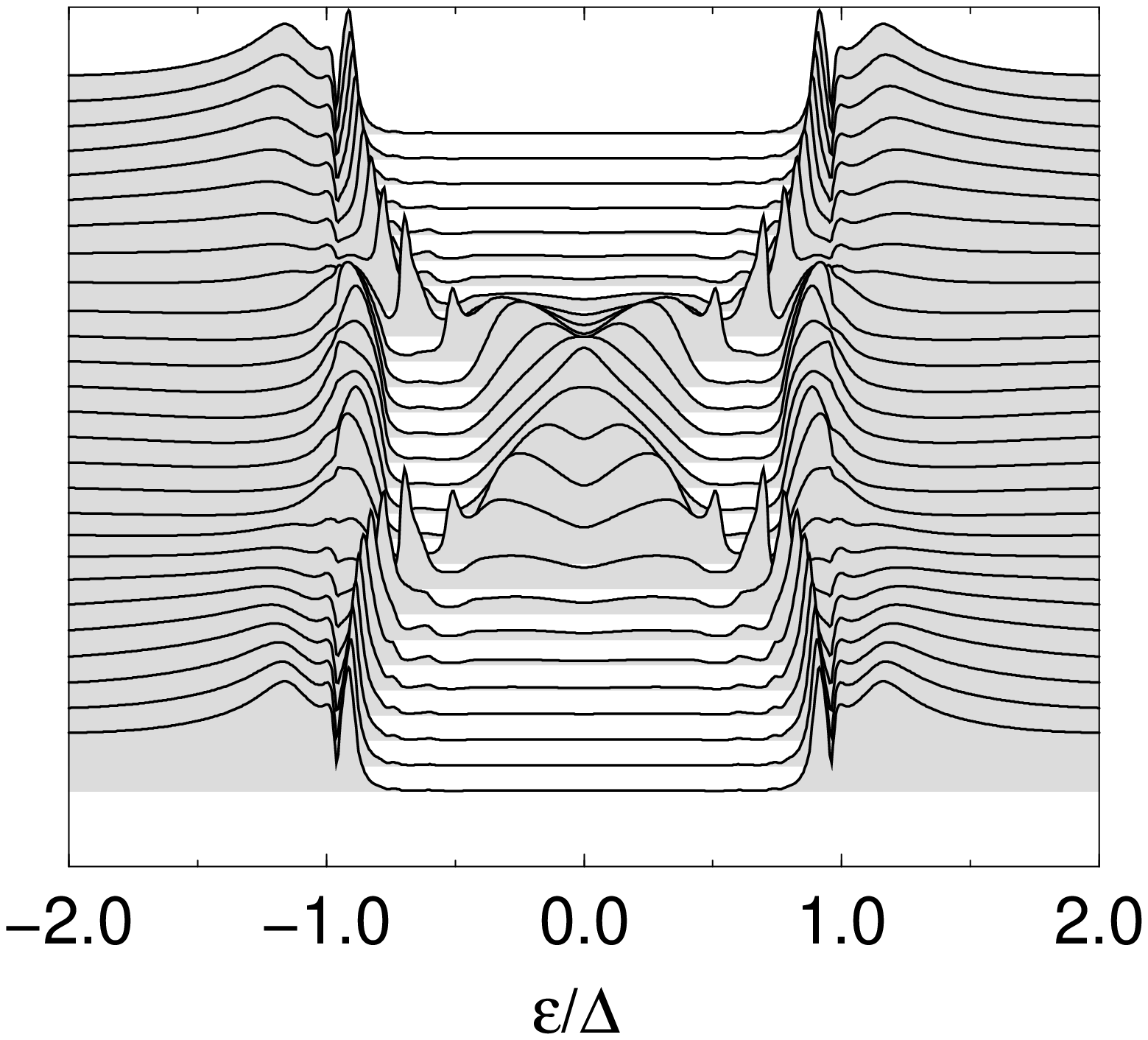}
\includegraphics[width=.49\hsize]{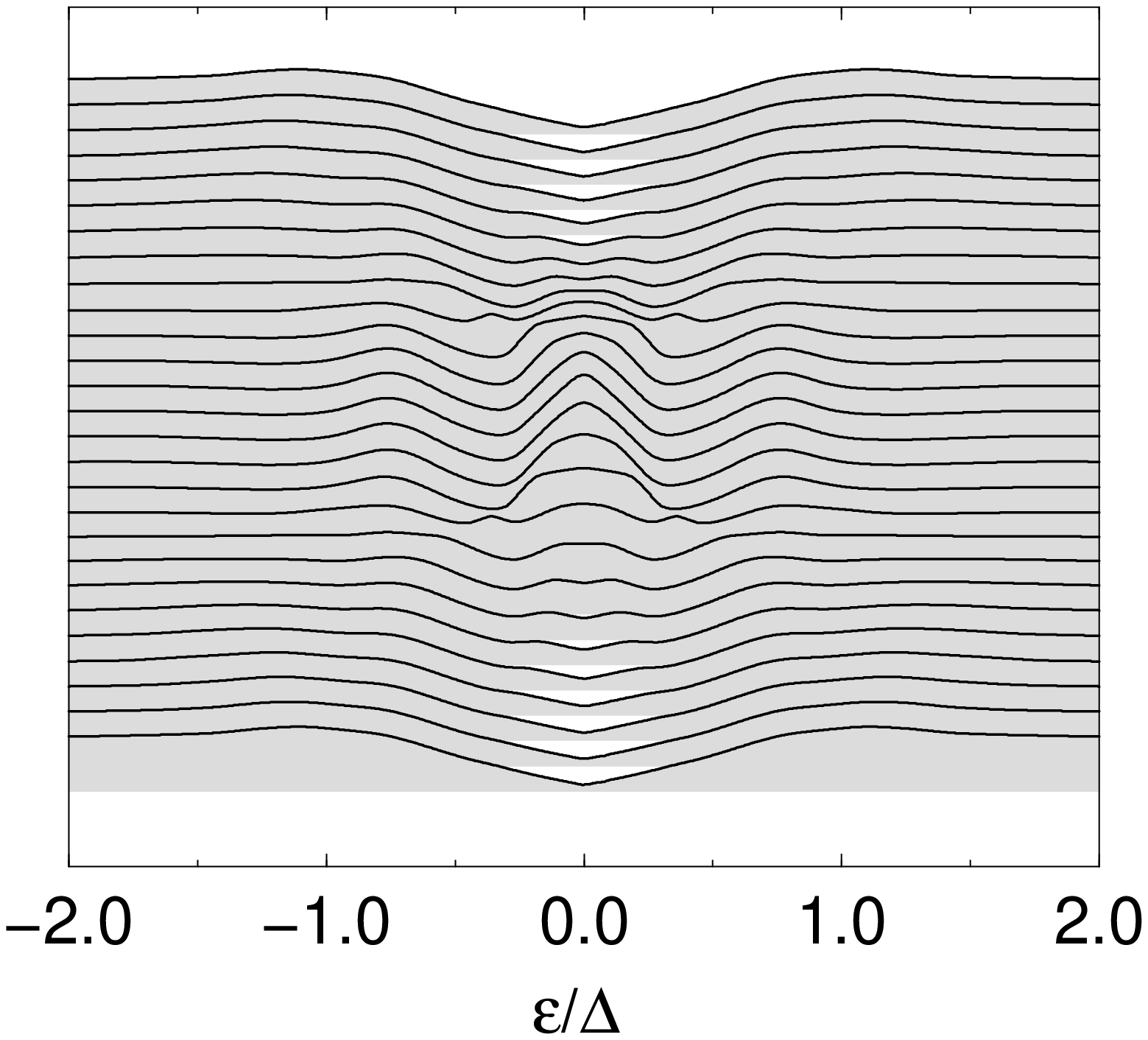}
\end{center}
\caption{ \label{DOS_pin_s}
Local density of states as a function of distance from the vortex center,
with a spacing of $\delta y = 0.79 \xi_0$, for a vortex pinned to an inclusion
of radius of $\xi_p=\pi\xi_0$. The left panel corresponds to $s$-wave symmetry
and the right panel is for $d$-wave symmetry along the anti-nodal direction.
The temperature is $T=0.3T_c$.
}
\end{figure}

Figures \ref{ADOS_pin_s} and \ref{DOS_pin_s} show the angle-resolved local
density of states and the LDOS for a vortex pinned on a metallic inclusion
of radius $\xi_{\mbox{\tiny pin}}=\pi\xi_0$. The order parameter, impurity
self energy and spectral densities were calculated self consistently for
impurity scattering in the Born limit. Qualitative changes resulting from
the inclusion occur inside the pinning center. The shape of the bound
state resonance lines are asymmetric in energy. The asymmetry arises
from multiple Andreev reflection
processes by the interface between the pinning center and the superconductor, 
which leads to additional bands of resonances
that overlap the vortex core resonances. 

\begin{figure}
\begin{center}
\leavevmode
\includegraphics[width=.63\hsize]{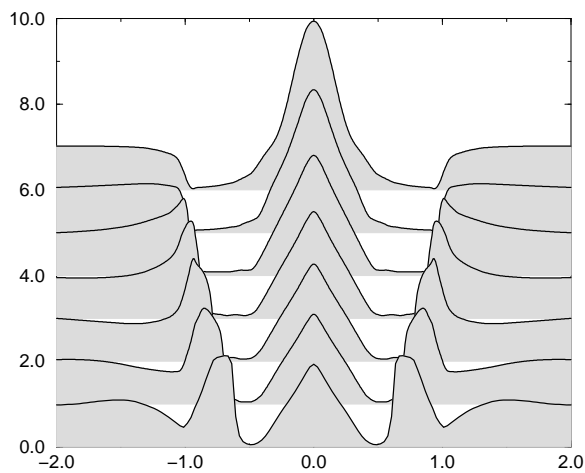}
\end{center}
\caption{ \label{pincompare}
Local density of states at the center of a vortex
pinned in the center of a normal inclusion. The spectra (from top to bottom)
correspond to different pinning radii: $\xi_{\mbox{\tiny pin}} = 0$, $0.79\xi_0$,
$1.57\xi_0$, $2.36\xi_0$, $3.14\xi_0$, $3.93\xi_0$, and $4.71\xi_0$.
The temperature is $T=0.3T_c$.
}
\end{figure}

In addition to the asymmetry in the linewidth of the resonances
the zero-energy bound state at the vortex center has a peculiar
spectral shape, shown in more detail in Fig. \ref{pincompare} for 
$s$-wave pairing symmetry, but also visible in the right panel of 
Fig. \ref{DOS_pin_s} for $d$-wave pairing symmetry as well.
In contrast to the spectra for a vortex without a pinning
center, the coherence peaks at the continuum edges are present at the vortex 
center.

Figure \ref{pincompare} shows the peaks near the 
continuum edges in the vortex center for a pinned vortex. The spectral 
weight near the continuum edge is taken from the resonance at the Fermi
level. The enhanced weight at the continuum edges is a precursor to
the formation of a secondary bound state that splits off from the continuum.
This can be seen in the evolution of the spectral density at the vortex
center as a function of the radius of the pinning center.
There is a zero energy resonance for all pinning 
radii, however, increasing the radius of the pinning center transfers
spectral weight from the zero-energy resonance to the continuum edge.
A coherence peak develops, splits off from the continuum edge, strengthens
and evolves to energies within the gap as the pinning radius changes from
$\xi_{\mbox{\tiny pin}} = 0.79\xi_0$ to $\xi_{\mbox{\tiny pin}} = 4.71\xi_0$.
Thus, the appearance of the coherence peak for pinning centers the
size of a coherence length or so is a precursor to the formation of a
secondary bound state within the gap. The spectral weight comes at the expense
of states just above the continuum edge and the zero-energy bound state,
which is diminished in intensity with increasing pinning radius.

\subsection{Doubly quantized vortices}

Vortices with winding numbers larger than $p = 1$
generally have line energies per unit winding number that are 
larger than that of a singly quantized vortex.\footnote{There are
counter examples for unconventional pairing with
a multi-component order parameter in which the lowest
energy vortex states are doubly quantized vortices \cite{tok90,mel92}.}
Nevertheless, doubly quantized vortices once formed are metastable
and in principle it should be possible to observe the rare doubly 
quantized vortex using an atomic probe such as a scanning tunneling
microscopic \cite{hes89f,ren91}. The spectrum of a doubly quantized
vortex differs in a fundamental way from that of a singly quantized
vortex. The singly quantized vortex, has a single branch of states
that disperse through zero energy at the vortex center. 
The zero mode is guaranteed in the quasiclassical
limit by $\pi$ change along trajectories that pass through the vortex
center. In contrast, there is no phase change along a trajectory through
the vortex center for a doubly quantized vortex, and thus no topological
requirement enforcing a zero energy bound state at the center of
a doubly quantized vortex. Nevertheless, there is a spectrum of 
bound states in the cores of doubly quantized vortices which lead
to characteristic structures in the LDOS and current spectral density 
of a doubly quantized vortex. 
\begin{figure}
\begin{center}
\leavevmode
\includegraphics[width=.49\hsize]{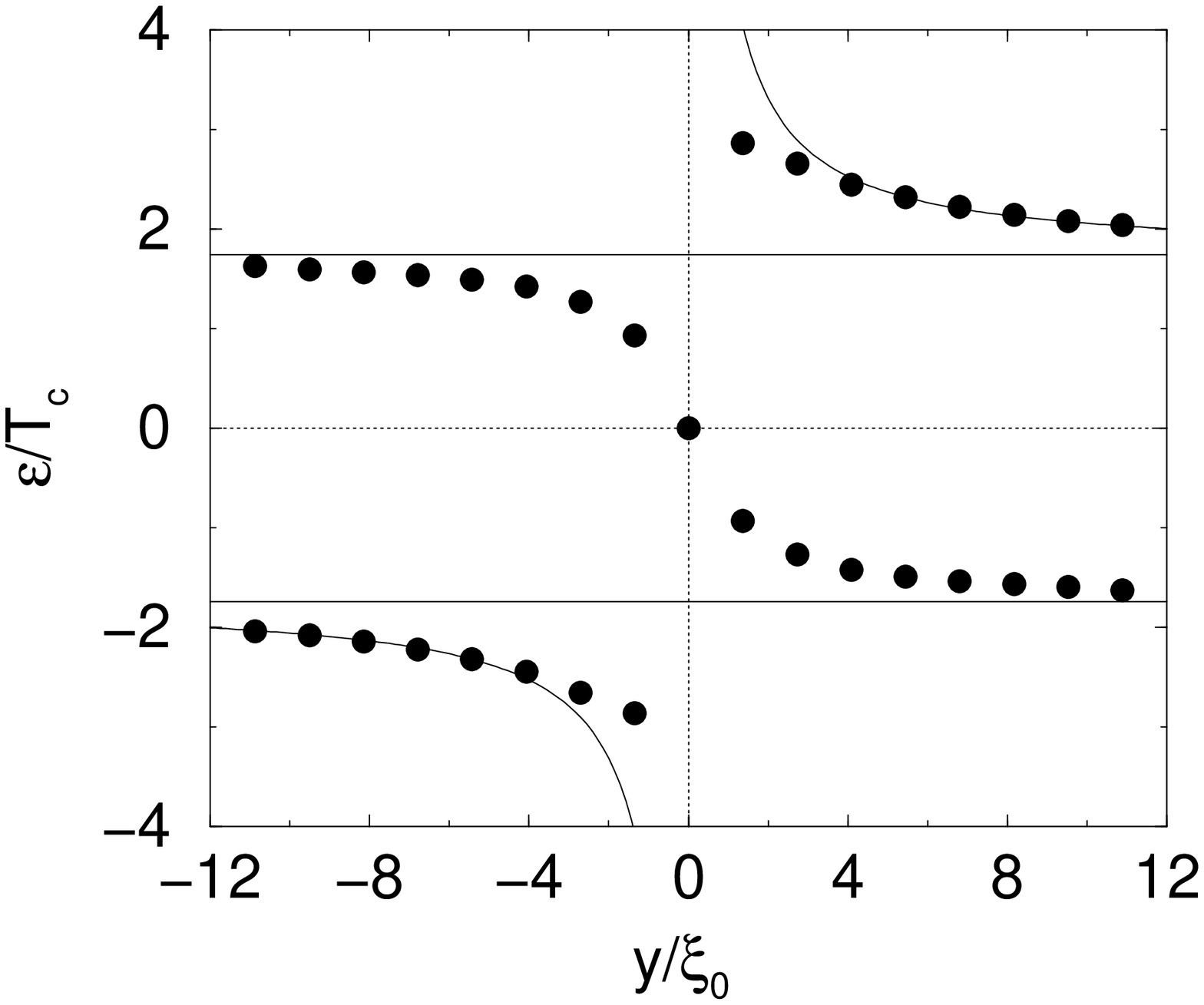}
\includegraphics[width=.49\hsize]{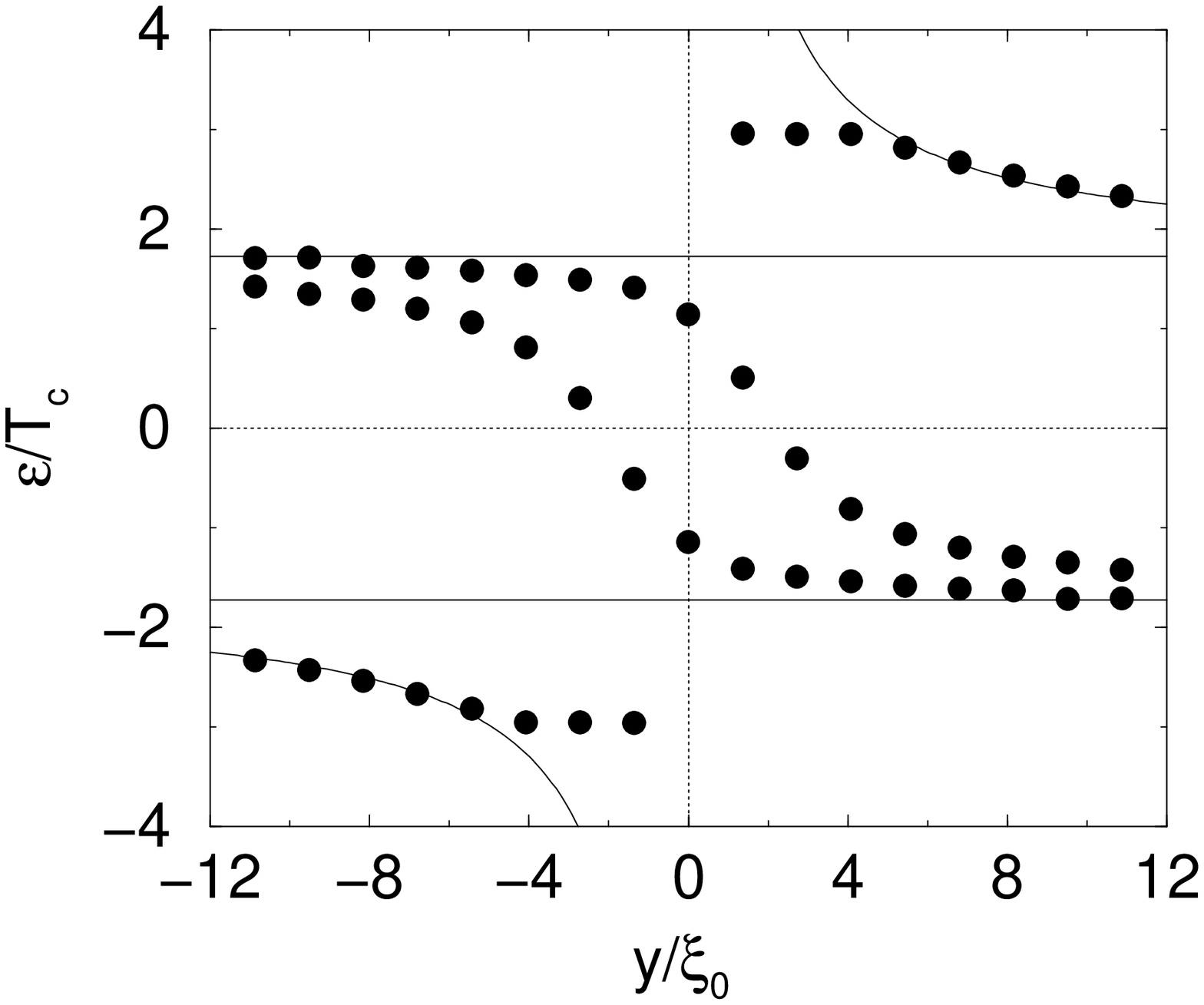}
\end{center}
\caption{ \label{bsdispersions}
Dispersion of the bound states (below the continuum edge) and of the coherence peaks
(above the continuum edge) for quasiparticle trajectories in $x$ direction
as a function of $y$, for a singly quantized vortex (left) and for a doubly
quantized vortex (right). The center of the vortex is at $y=0$. Born impurity
scattering with a mean free path of $\ell =10\xi_0$ is assumed.
The temperature is $T=0.3T_c$. The thin lines show the energies 
of the continuum edges and the Doppler shifted energies of the coherence
peaks, assuming the London form for the condensate momentum
$\vec{p}_s= p\vec{e}_{\phi}/2r$, where $p$ is the winding number.
The dispersion is shown for s-wave pairing; the corresponding
data for d-wave pairing is similar.
}
\end{figure}

\begin{figure}
\begin{center}
\leavevmode
\includegraphics[width=.49\hsize]{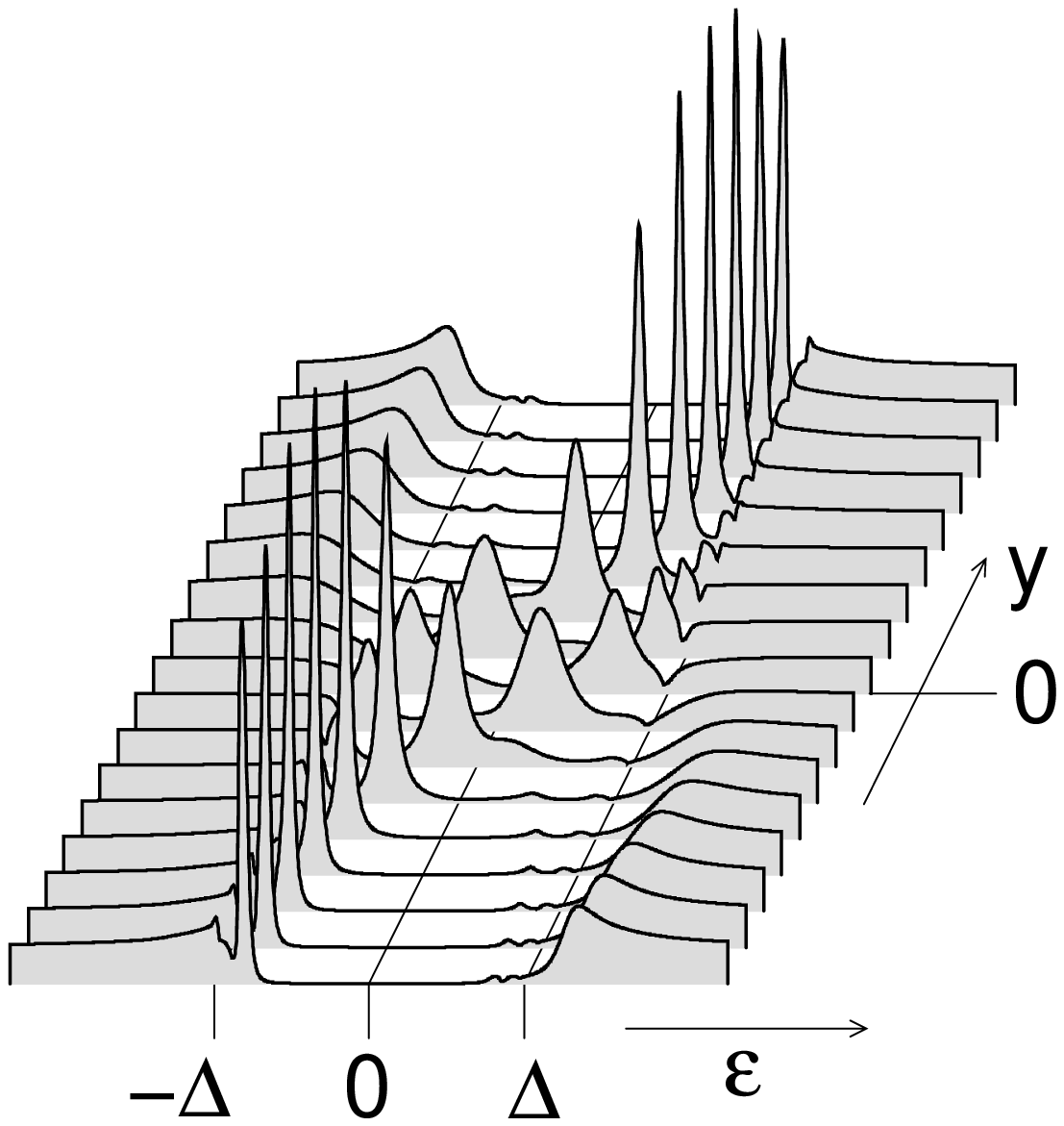}
\includegraphics[width=.49\hsize]{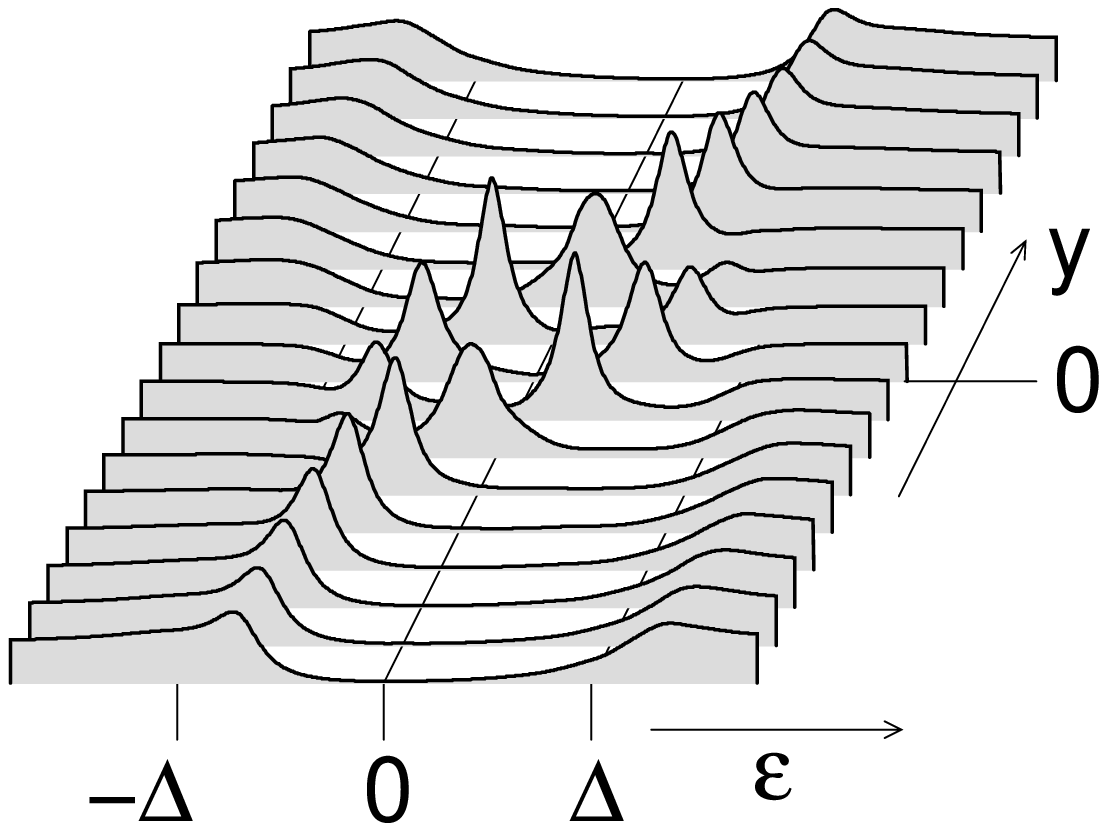}
\end{center}
\caption{ \label{Double_ADOS}
Angle resolved density of states for a doubly quantized vortex
along quasiparticle trajectories parallel to the
$x$-direction, as a function of impact parameter, $y$, with a spacing 
$\delta y = 1.36\xi_0$.
The left panel corresponds to $s$-wave symmetry and the right panel 
is for $d$-wave symmetry. Impurity scattering is included self-consistently
in the Born limit with a mean free path $\ell = 10 \xi_0$.
The temperature is $T=0.3T_c$.
}
\end{figure}

This structure was discussed for a doubly quantized vortex in the 
superclean limit in Ref. \cite{rai96}. Figure \ref{Double_ADOS} shows 
the angle-resolved density of states for 
a doubly quantized vortex in an $s$-wave superconductor for
trajectories parallel to the $x$-axis at different impact parameters along
the $y$-direction. {\sl Two} branches of vortex bound states cross the 
Fermi level at distances of order a coherence length from the vortex center.
Thus, zero-energy bound states exist in the core but they are localized
(for $s$-wave pairing symmetry) on trajectories at finite impact parameter
from the vortex center. The locus of these trajectories forms a circle
of radius $r_{\mbox{\tiny bs}}\simeq 2.5\xi_0$ around the vortex center.
Also note that the Doppler shift of the continuum spectrum is twice that 
for singly quantized vortices.

This can be seen by comparing the spectra 
far from the core in Figs. \ref{Double_ADOS} and \ref{Spec1a}.
The distinctive features in the spectrum of a doubly quantized vortex
are present for either $s$- and $d$-wave pairing symmetry.
The doubling of bound state branches that cross the Fermi level is observed
for both pairing symmetries. The main difference in the spectra,
as in the case of singly quantized vortices, is in the width of 
the resonances.

\begin{figure}
\begin{center}
\leavevmode
\includegraphics[width=.49\hsize]{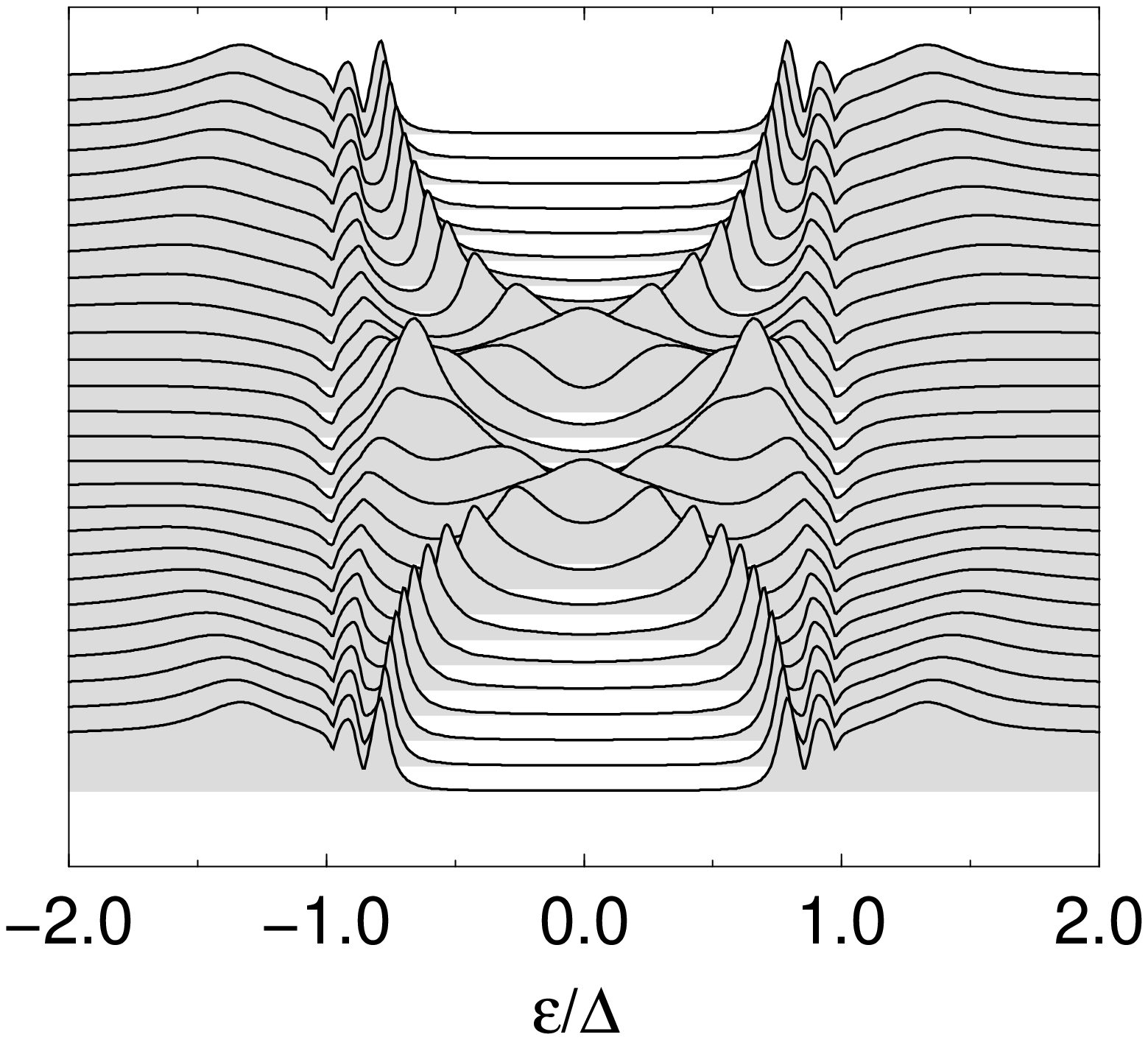}
\includegraphics[width=.49\hsize]{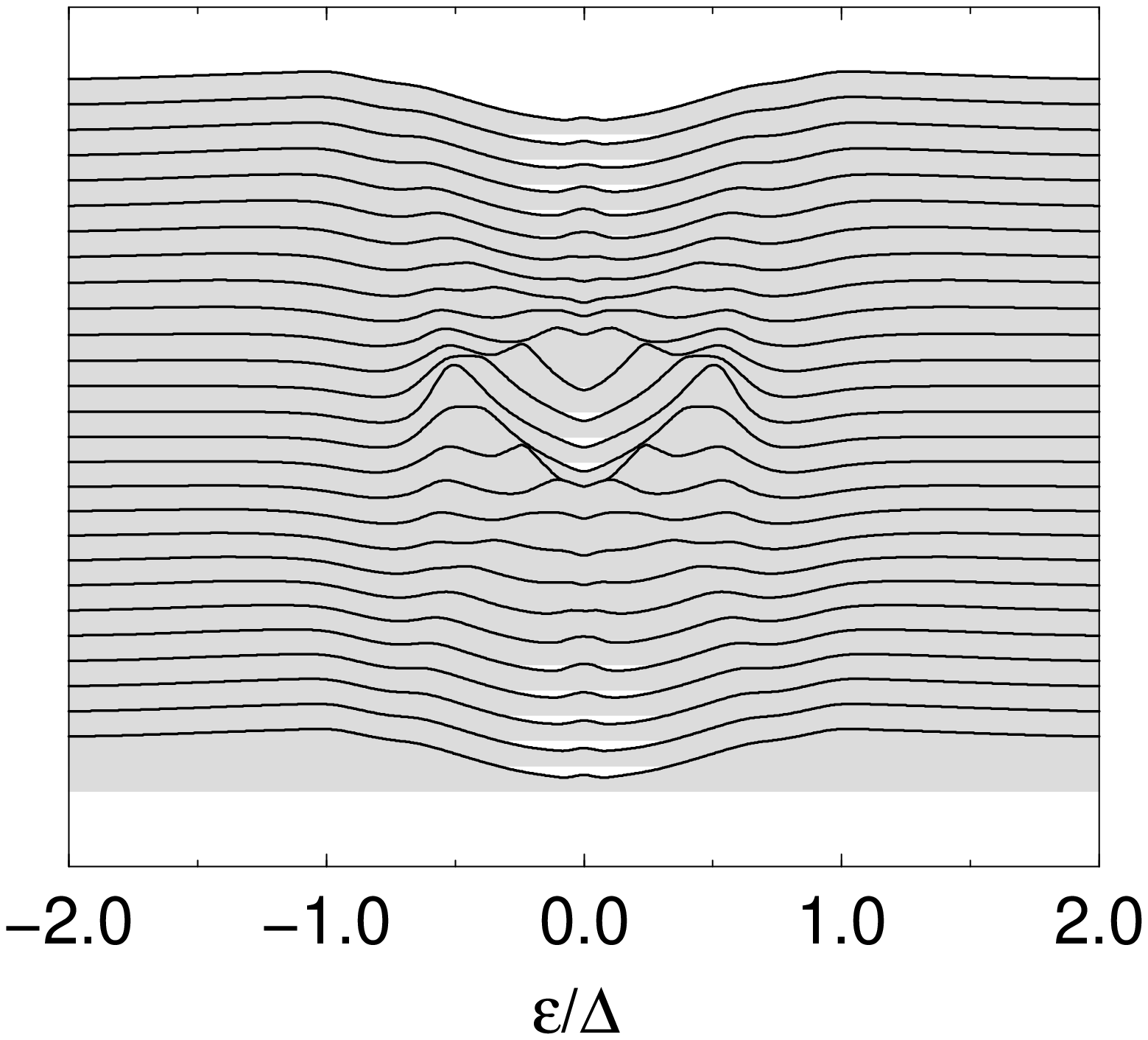}
\end{center}
\caption{\label{Double_DOS}
Local density of states 
for doubly quantized vortices along trajectories 
parallel to the $x$-axis as a function of impact parameter along the 
$y$-axis with a spacing of $\delta y =1.36\xi_0$. The left panel
corresponds to $s$-wave pairing symmetry, and the right panel
is for $d$-wave pairing along the anti-nodal direction.
Impurity scattering is included in the 
Born limit with a mean free path of $\ell = 10 \xi_0$ and the
temperature is $T=0.3T_c$.
}
\end{figure}

\begin{figure}
\begin{center}
\leavevmode
\includegraphics[width=.49\hsize]{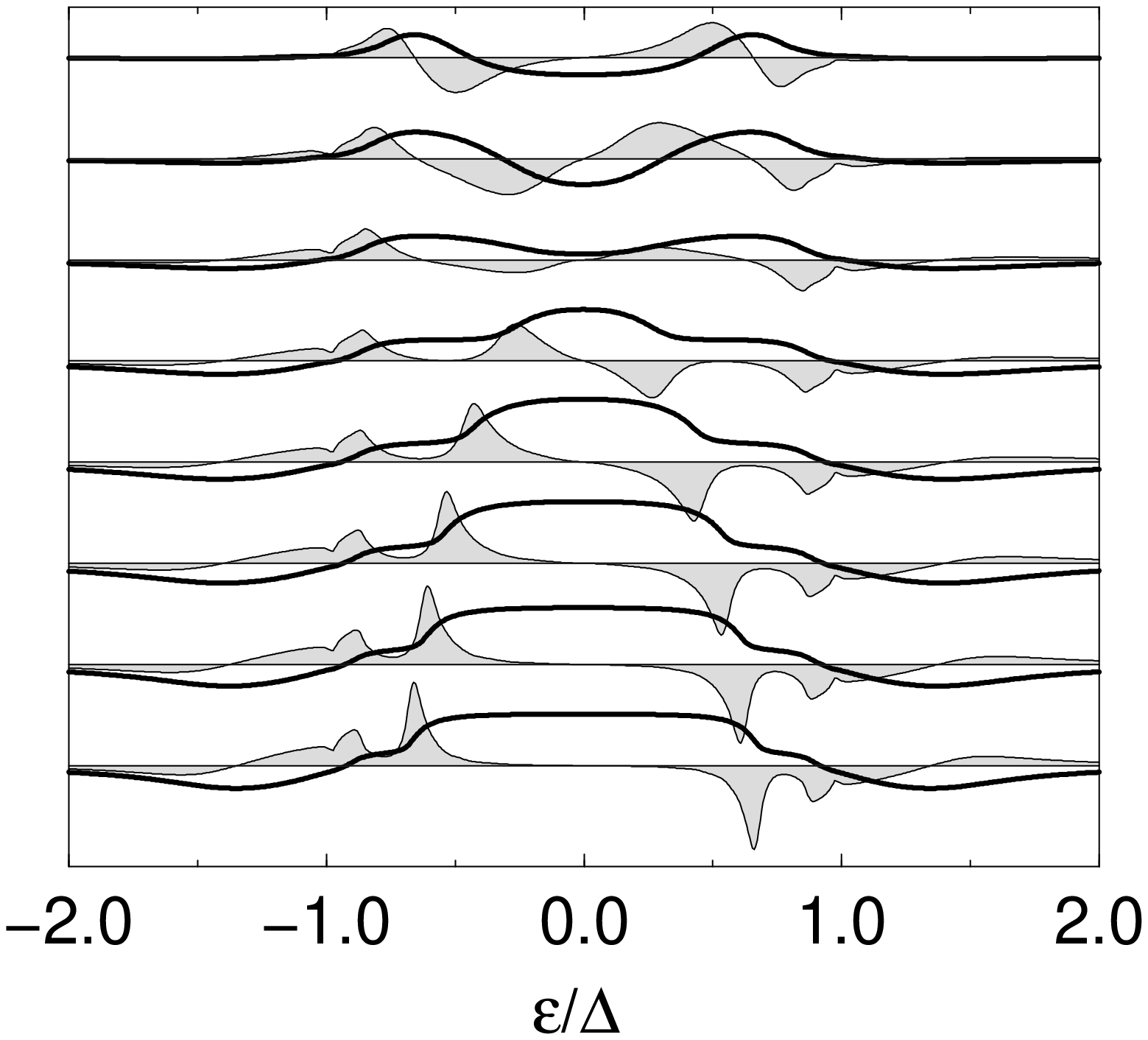}
\includegraphics[width=.49\hsize]{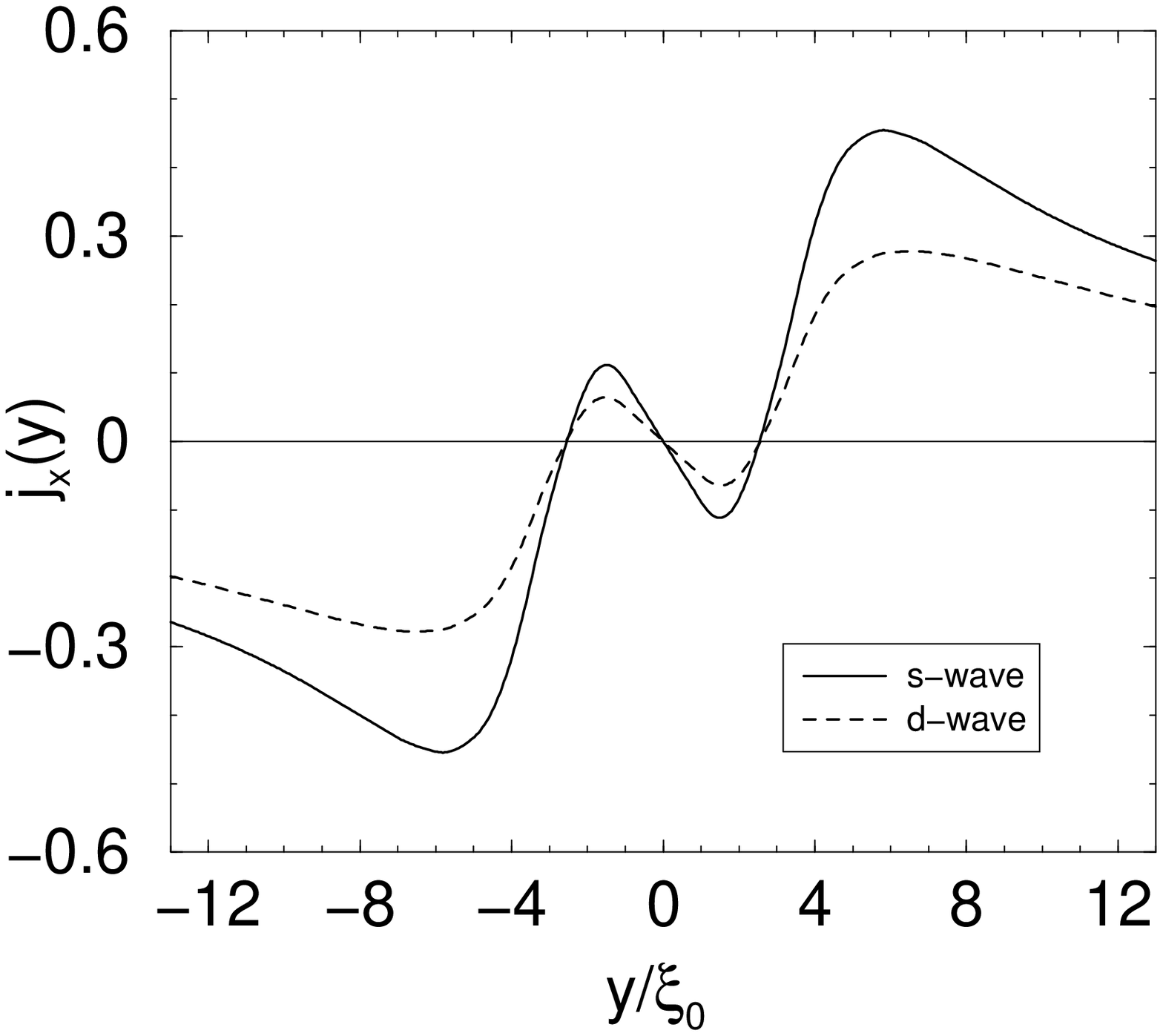}
\end{center}
\caption{\label{Current_Reversal}
The spectral current density for a doubly quantized vortex is shown 
along a trajectory parallel to the $x$-axis as a function of impact 
parameter ($y$-axis) with a spacing $\delta y =1.36\xi_0$, starting 
at $y =1.36\xi_0$. The thick curve is the cumulative spectral current
density as a function of $\epsilon$, obtained by integrating $j_x(\epsilon;y)$
from $-\infty$ to $\epsilon$.
The right panel is the x-component of the current density 
as a function of impact parameter for $s$-wave (solid line) and 
d-wave (dashed line) pairing.
}
\end{figure}

The most distinguishing feature of doubly quantized vortices is that
the supercurrents near the center of the vortex flow {\sl counter} to the 
asymptotic superflow associated with the phase winding 
around the vortex \cite{rai96}. The counter circulating 
currents in the core, shown in the right panel of 
Fig. \ref{Current_Reversal}, are due to the bound states
interior to the radius defined by the zero-energy bound state
(Note that the zero-energy bound state itself does not carry current). 
This structure is revealed in 
Figs. \ref{Double_DOS} and \ref{Current_Reversal},
which show the LDOS and spectral current density for
a doubly quantized vortex with $s$-wave 
and $d$-wave pairing symmetry.
The left panel of Fig. \ref{Current_Reversal} also shows the cumulative
spectral current density and the reversal of the current for 
as one branch of bound states disperses below
the Fermi level for $r<r_{\mbox{\tiny bs}}$.

These spectra show that the current density near the vortex
core is carried mainly by the bound states, and that the reversal of
the current direction near the vortex center is due to the 
branch of counter-flowing bound states dispersing below the Fermi
level for impact parameter $r < r_{\mbox{\tiny bs}}$.
In this region of the core the co-moving bound state is above
the Fermi level so the asymmetry in the level occupation
produces a counter flowing current.

\section{Nonequilibrium Response}\label{Sec:Nonequilibrium_Response}

The dynamics of the electronic excitations of the 
vortex core play a key role in the dissipative processes in 
type II superconductors. Except in the dirty limit, $\ell \ll \xi_0$,
the response of the core states to an electromagnetic field is 
generally very different from that of normal electrons.
It is energetically unfavorable to maintain a
charge density of the order of an elementary charge over
a region with diameter of order the coherence length.
Instead an electrochemical potential is induced which
ensures that almost no net charge accumulates in the core 
region. However, a dipolar-like charge distribution develops
which generates an internal electric field in the core.
The internal field varies on the scale of the coherence
length. This leads to a nonlocal response of the quasiparticles
to the total electric field, even when the applied field varies 
on a much longer length scale and can considered to be homogeneous.
The dynamical response of the vortex core includes the collective
mode of the inhomogeneous order parameter.
This mode couples to the electro-chemical
potential, $\delta\Phi$, in the vortex core region. 
This potential is generated by
the charge dynamics of vortex core states and gives rise to
internal electric fields which in turn drive the current density 
and the order parameter near the vortex core region.
The induced electric fields in the core are the same order
of magnitude as the external field. The dynamics of the 
core states are strongly coupled to the charge 
current and collective mode
of the order parameter. Thus, the determination of the induced 
order parameter, as well as the spectrum and distribution
function for the core states and non-equilibrium impurity scattering 
processes requires dynamical self consistency.
Numerous calculations of the a.c. response neglect the self-consistent
coupling of the collective mode and the spectral dynamics, or concentrate
on the $\omega\rightarrow 0$ limit \cite{jan92,hsu95,kop78a,kop95}.
Quasiclassical theory is the only formulation of the theory of 
nonequilibrium superconductivity presently
capable of describing the nonlocal response of the order parameter
and quasiparticle dynamics in the presence of mesoscopic 
inhomogeneities and disorder.
The numerical solution to the self-consistency problem was presented for 
unpinned vortices in Ref. \cite{esc99}. Here we report results for the
the electromagnetic response of isolated vortices bound to a pinning
center in a superconductor with $s$-wave pairing symmetry.

\subsection{Dynamical charge response}
\label{chargeresp}

The charge dynamics of layered superconductors has two distinct origins.
The c-axis dynamics is determined by the Josephson coupling between
the conducting planes. Here we are concerned with the in-plane 
electrodynamics associated with the response of the order parameter 
and quasiparticle states bound to the vortex core. We assume strong 
Josephson coupling between different layers, and neglect variations 
of the response between different layers. This requires that the 
polarization of the electric field be in-plane, so that there is no 
coupling of the in-plane dynamics to the Josephson plasma modes.
The external electromagnetic field is assumed to be long wavelength
compared to the size of the vortex core, $\lambda_{\mbox{\tiny EM}}\gg\xi_0$.
In this limit we can assume the a.c. electric field to be uniform and described 
by a vector potential, $\vE_{\omega}(t)=-\frac{1}{c}\partial_t\vA_{\omega}$.
We can also neglect the response to the a.c. magnetic field
in the limit $\lambda\gg\xi_0$. In this case the spatial variation of the 
induced electric field occurs mainly within each conducting layer
on the scale of the coherence length, $\xi_0$. Poisson's equation 
implies that induced charge densities are of order $\delta\Phi/\xi_0^2$,
where $\delta\Phi$ is the induced electrochemical potential in the core.
This leads to a dynamical charge of order $e\,(\Delta/E_f)$ in the vortex core.
Once the electrochemical potential is calculated from 
Eq. (\ref{resplocalneutral}) we can calculate the charge density 
fluctuations of order $(\Delta/E_f)^3$ from Poisson's equation,
\be
\rho^{(3)}(\vec{R};t)=-\frac{1}{4\pi}\grad^2\,\Phi(\vec{R};t)
\,.
\ee

\begin{figure}
\begin{center}
\leavevmode
\includegraphics[width=.49\hsize]{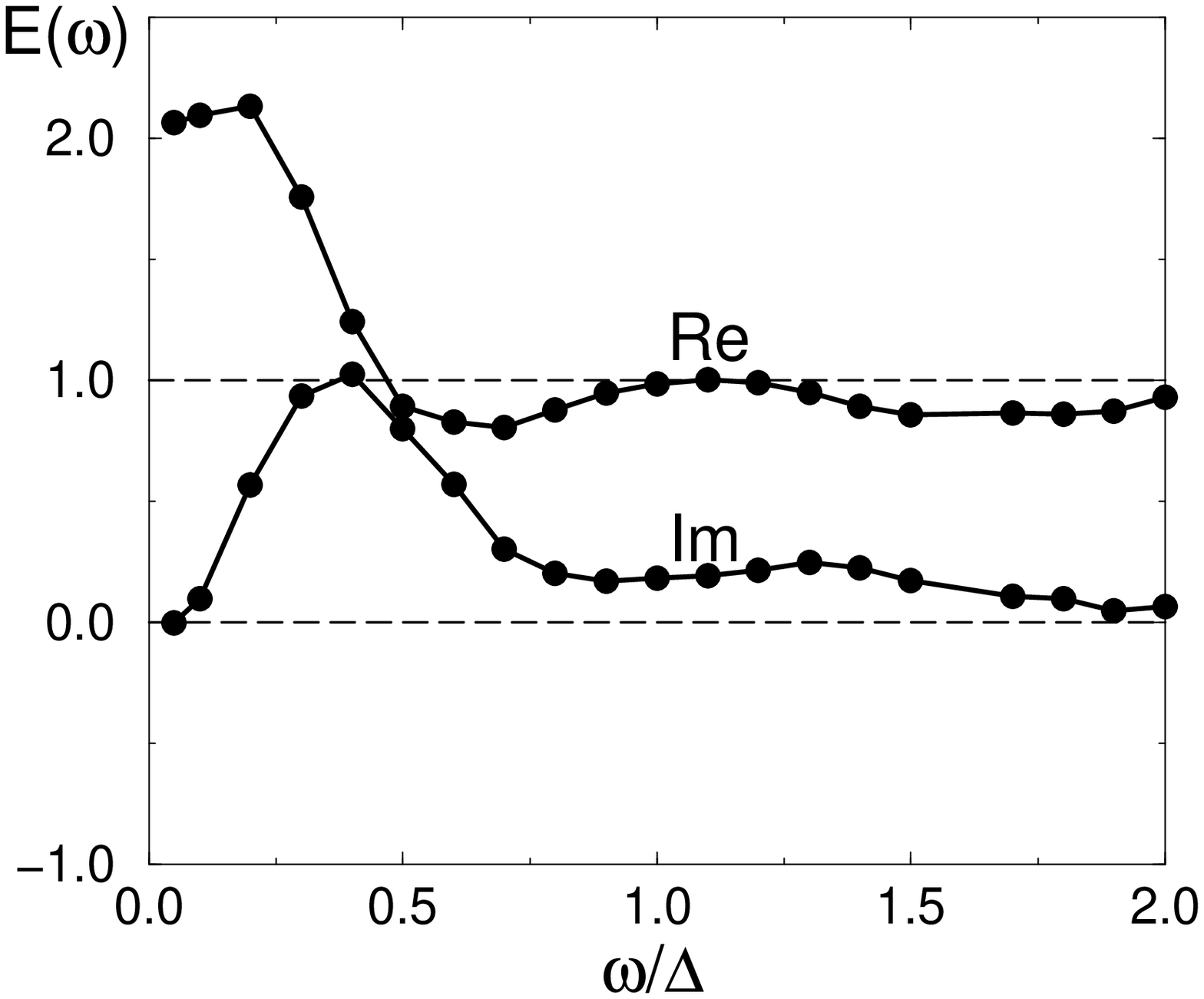}
\includegraphics[width=.49\hsize]{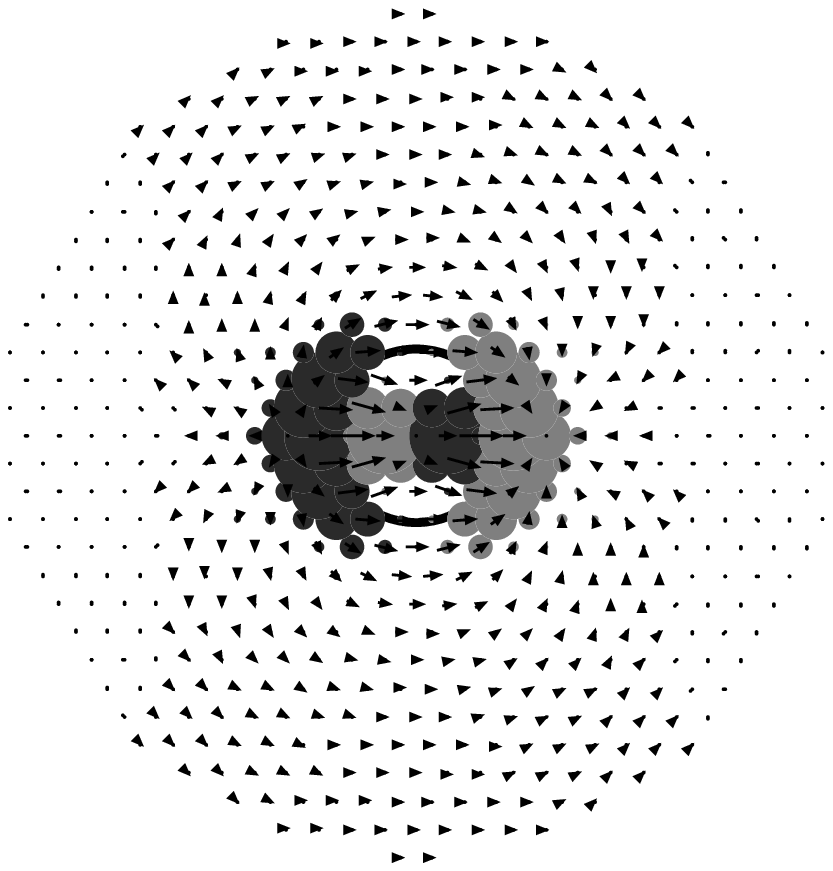}
\end{center}
\caption{ \label{FELD}
Left: Total electric field in the center of a pinned $s$-wave vortex
as a function of frequency $\omega $ of an external
a.c. electric field with polarization vector in the $x$-direction and
wavelength large compared to $\xi_0$.
Right: The corresponding in phase charge response around the pinning site
for frequency $\omega = 0.1 \Delta $. Gray corresponds to negative charge
and black to positive charge. 
The arrows denote the total electric field vectors.
The pinning center is a circular normal metallic inclusion with a radius
$\xi_p=1.57 \xi_0$, shown as the black circle.
Impurity scattering is included in the Born limit with a 
mean free path of $\ell=10\xi_0$. The temperature is $T=0.3T_c$ and
the calculations are carried out in the high-$\kappa$ limit.
}
\end{figure}

In Fig. \ref{FELD} we show the total electric field (external plus
induced) acting on the quasiparticles in units of the external field.
For $\omega\gsim 2\Delta$ the total field is approximately equal to the
external field. However, at frequencies $\omega < 2\Delta$ an
out-of-phase response develops. For $\omega\ll\Delta$, the total field 
in the pinning region approaches twice the external field, and
the out-of-phase component vanishes. In the intermediate frequency region,
$0.1\Delta\lsim\omega\lsim\Delta$, both in-phase and out-of-phase 
components are comparable. The right panel of Fig. \ref{FELD} shows 
the charge distribution for $\omega=0.1\Delta$ which oscillates out of phase 
with the external field. A dipolar charge distribution accumulates 
at the interface between the superconductor and the normal inclusion, 
oscillating at the frequency of the external field. At the center of the 
pinning site the out-of-phase component of the field is nearly zero
at low frequencies (see also the left panel).
The induced charge which accumulates is of order of $e\Delta/E_f$ within
a region of order $\xi_0^2$ in each conducting layer. 
This charge is a factor of 
$E_f/\Delta$ larger than the static charge of a vortex that arises from
particle-hole asymmetry \cite{kho95,fei95,bla96}.

\subsection{Local Dynamical Conductivity}

Because the total electric field varies on the scale of
a coherence length, the current response expressed in terms of the 
the total field is nonlocal in the intermediate clean regime.
However, we can define a local conductivity tensor in terms of the
response to the external field, provided the external field varies 
on a length scale large compared to the coherence length,
\be
J_{\mu}(\vec{R},\omega) =
\sigma_{\mu\nu}(\vec{R},\omega)E^{\mbox{\tiny ext}}_{\nu}(\omega)
\,.
\ee

\begin{figure}
\begin{center}
\leavevmode
\includegraphics[width=.49\hsize]{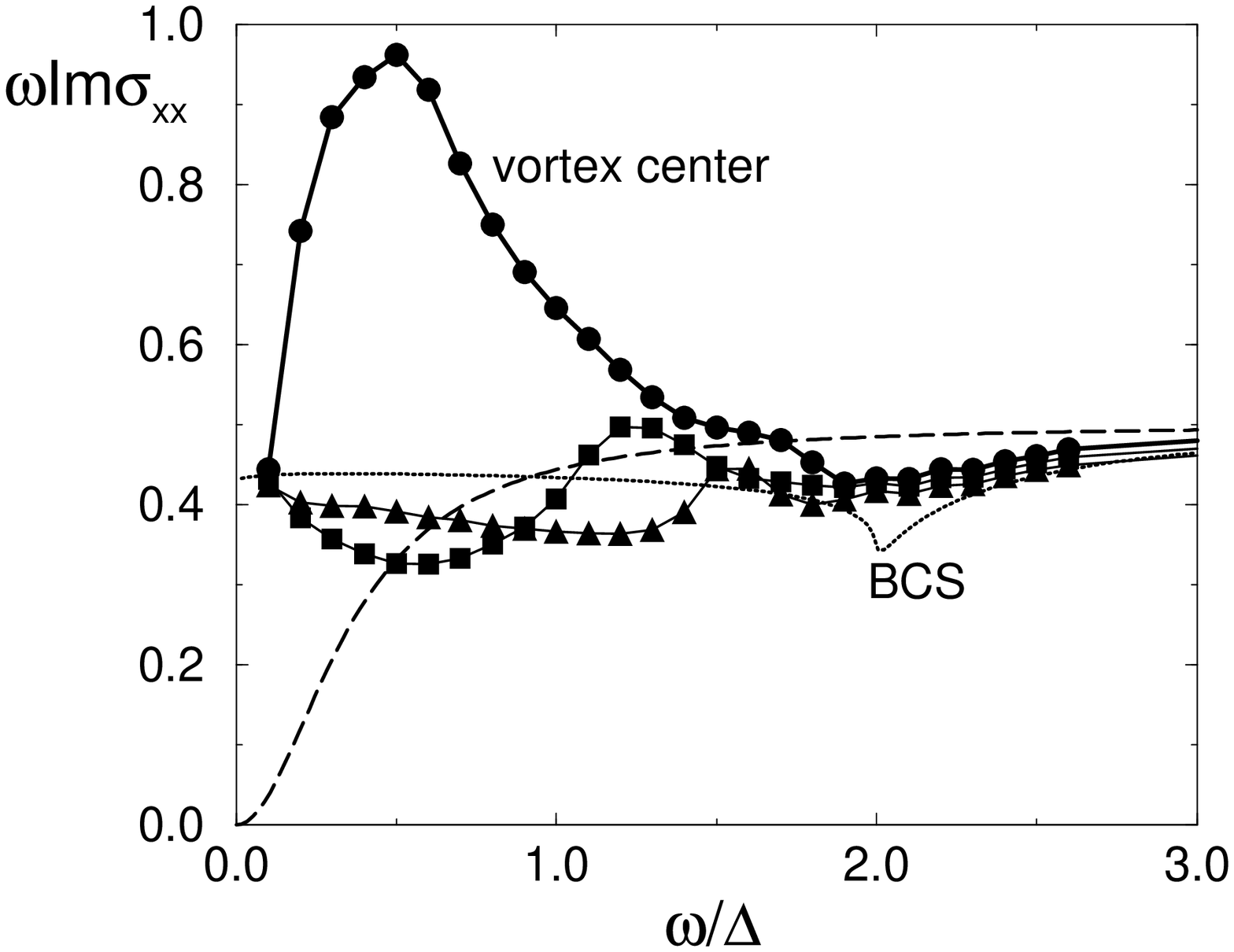}
\includegraphics[width=.49\hsize]{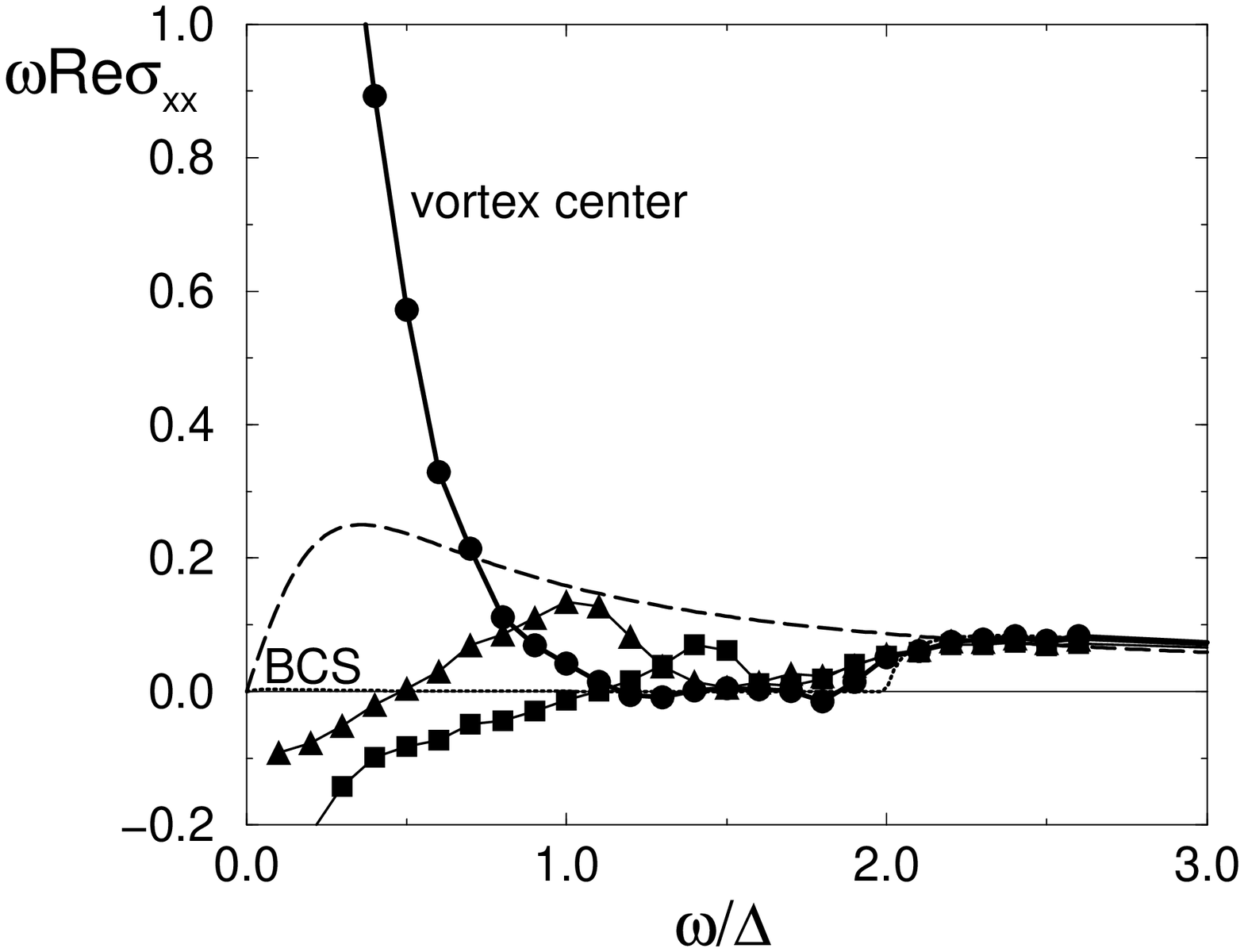}
\end{center}
\caption{ \label{cond_unpin}
Imaginary (left) and real part (right) of the local conductivity
(multiplied with $\omega $ for convenience) 
in the center of an unpinned $s$-wave vortex
(circles), and at different distances from the center
along the $y$-axis: $1.1 \xi_0$ (squares), and $2.2 \xi_0$ (triangles),
as a function of frequency $\omega $ of an external
a.c. electric field with polarization vector in $x$-direction and
wavelength large compared to $\xi_0$.
Impurity scattering is taken into account 
in Born limit with a mean free path of $\ell=10\xi_0$.
The temperature is $T=0.3T_c$. Calculations are done in the 
high-$\kappa $ limit.
Also shown are the response for a homogeneous $s$-wave superconductor
(dotted, denoted `BCS'), and the Drude conductivity of the normal 
metal (dashed).
}
\end{figure}

\begin{figure}
\begin{center}
\leavevmode
\includegraphics[width=.49\hsize]{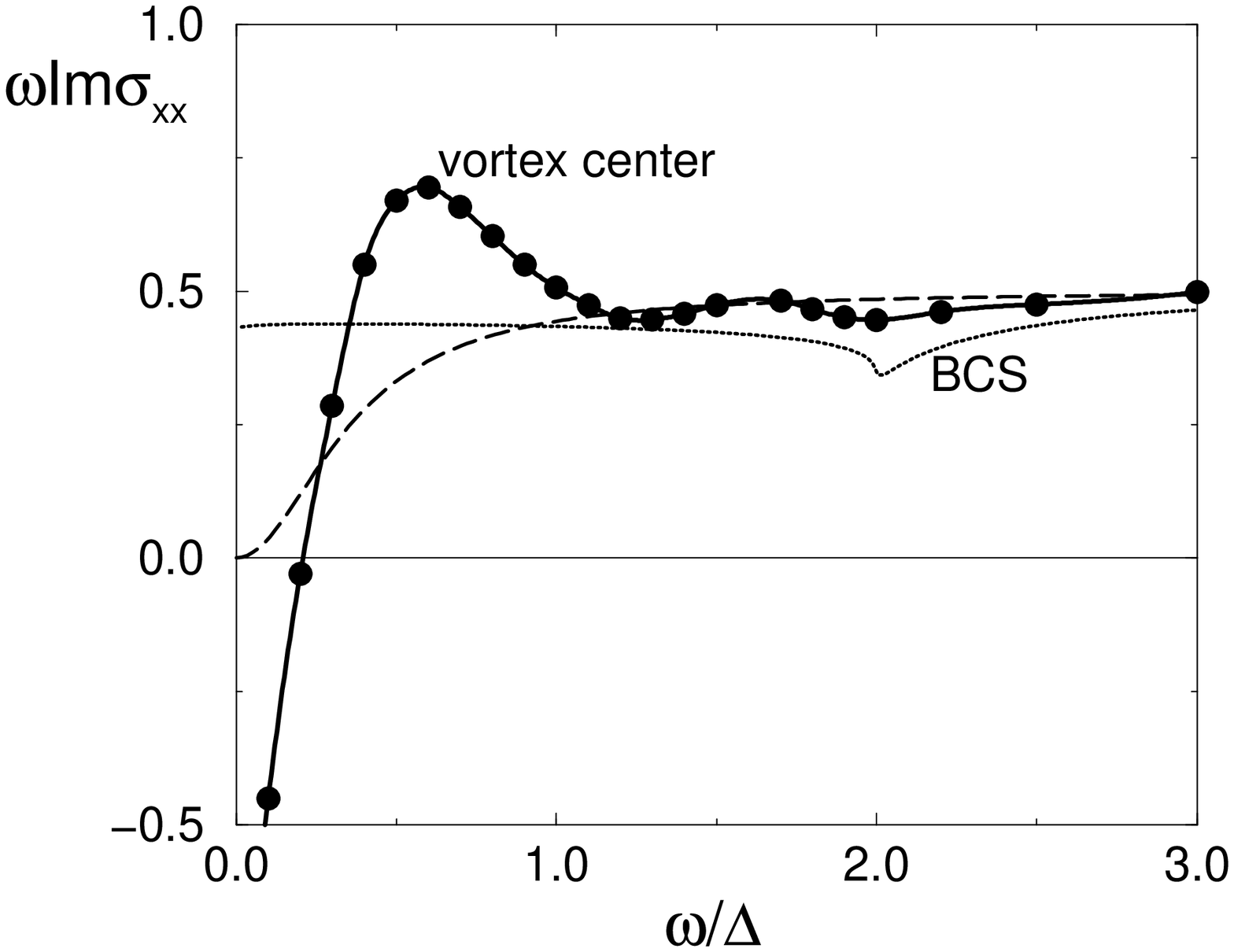}
\includegraphics[width=.49\hsize]{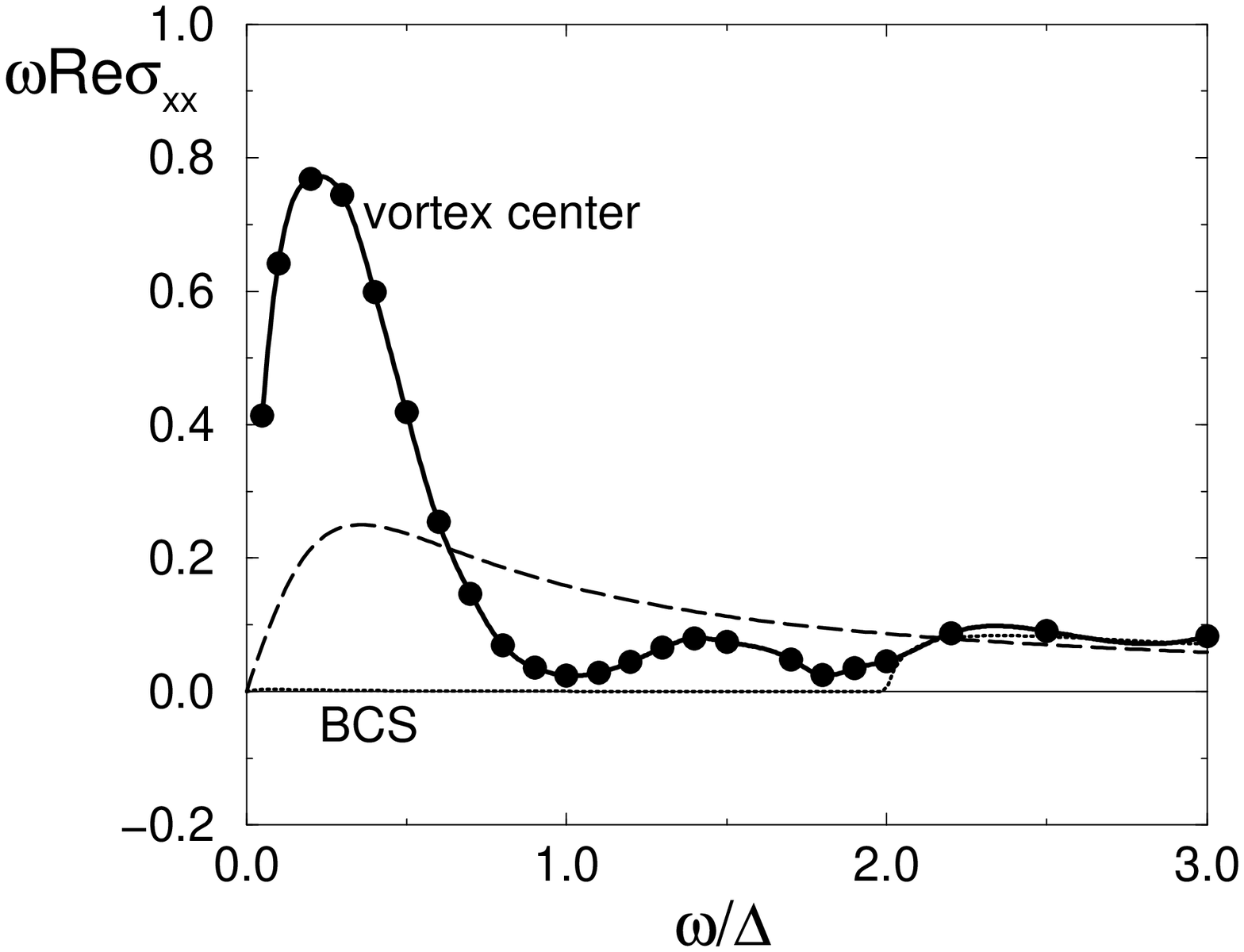}
\end{center}
\caption{ \label{cond_pin}
Like Fig. \ref{cond_unpin} for a pinned vortex with a pinning radius
$\xi_p=1.57 \xi_0$. As pinning site a circular normal conducting
inclusion located at the center of the vortex was assumed 
(see Fig. \ref{FELD}).
}
\end{figure}

Figures \ref{cond_unpin} and \ref{cond_pin} show results
for the conductivity, $\sigma_{||}$, in the vortex core 
region as a function of frequency for both unpinned and pinned 
vortices. For the pinned vortex the radius of the pinning
center is $\xi_{\mbox{\tiny pin}}=1.57\xi_0$.

First consider the unpinned vortex. The absorptive part 
of the conductivity (right panel of Fig. \ref{cond_unpin})
is strongly enhanced
at the vortex center compared to the normal-state Drude 
conductivity. The reactive response exhibits a maximum 
at a frequency determined by the impurity scattering rate. 
A few coherence lengths away from the vortex center the real part of 
the conductivity changes sign at low frequencies. This is a
signature that energy is transported by vortex-core excitations
away from the vortex center producing ``hot spots'' outside
the core. The net dissipation is determined by 
inelastic scattering processes in the region
around the vortex core. At distances of order a coherence length 
or so from the vortex center there is also structure 
in the conductivity spectrum at higher frequencies
reflecting absorptive transitions between quasiparticle 
excitations with energies corresponding to the Van Hove 
singularities in the local density of states. The maxima in the 
absorptive part of the conductivity at $y=1.36\xi_0$ and $2.72\xi_0$
from the center correspond to the energy level separation between
the Van Hove peaks above and below the Fermi level shown
in Fig. \ref{Nofepss}.

The conductivity spectra for a vortex pinned by a metallic inclusion
at its center is shown in Fig. \ref{cond_pin}. 
The most significant difference compared to the unpinned vortex
occurs at frequencies $\omega < 1/\tau$. The absorptive part
of the conductivity (right panel) is reduced compared to that of the unpinned 
case at low frequencies. The three broad peaks in the absorption
spectrum correspond to scattering and dissipation within the zero-energy 
resonance (the dominant low-frequency peak), transitions between
the zero-energy resonance and the continuum (the peak near 
$\omega\sim 1.5\Delta$), and pair-breaking transitions from the negative 
energy to positive energy continuum states (broad peak above $2\Delta$).
The other notable feature is the reactive response at low frequencies
which becomes {\sl negative} in the low frequency limit, corresponding 
to superflow in the core that is counter to the induced supercurrent 
outside the vortex core and pinning center. This counterflow is
required in order to satisfy the conductivity sum rule.
The counterflow in the vortex center is also present for unpinned
vortices, but at smaller frequencies for this particular
impurity scattering rate. The low-frequency counterflow is also
related to characteristic current patterns associated with low-frequency
vortex dynamics which we discuss below.

\begin{figure}
\begin{center}
\leavevmode
\includegraphics[width=.49\hsize]{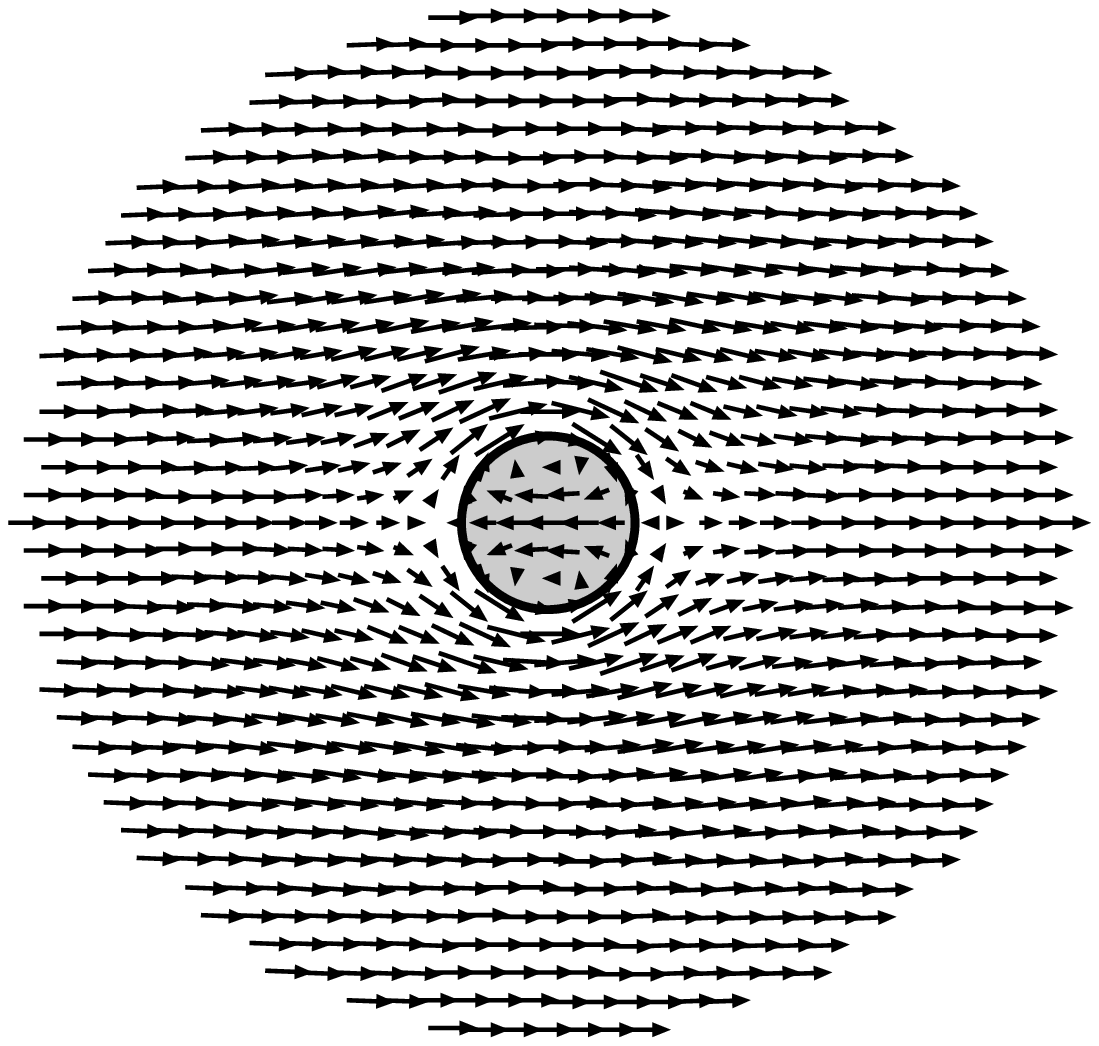}
\includegraphics[width=.49\hsize]{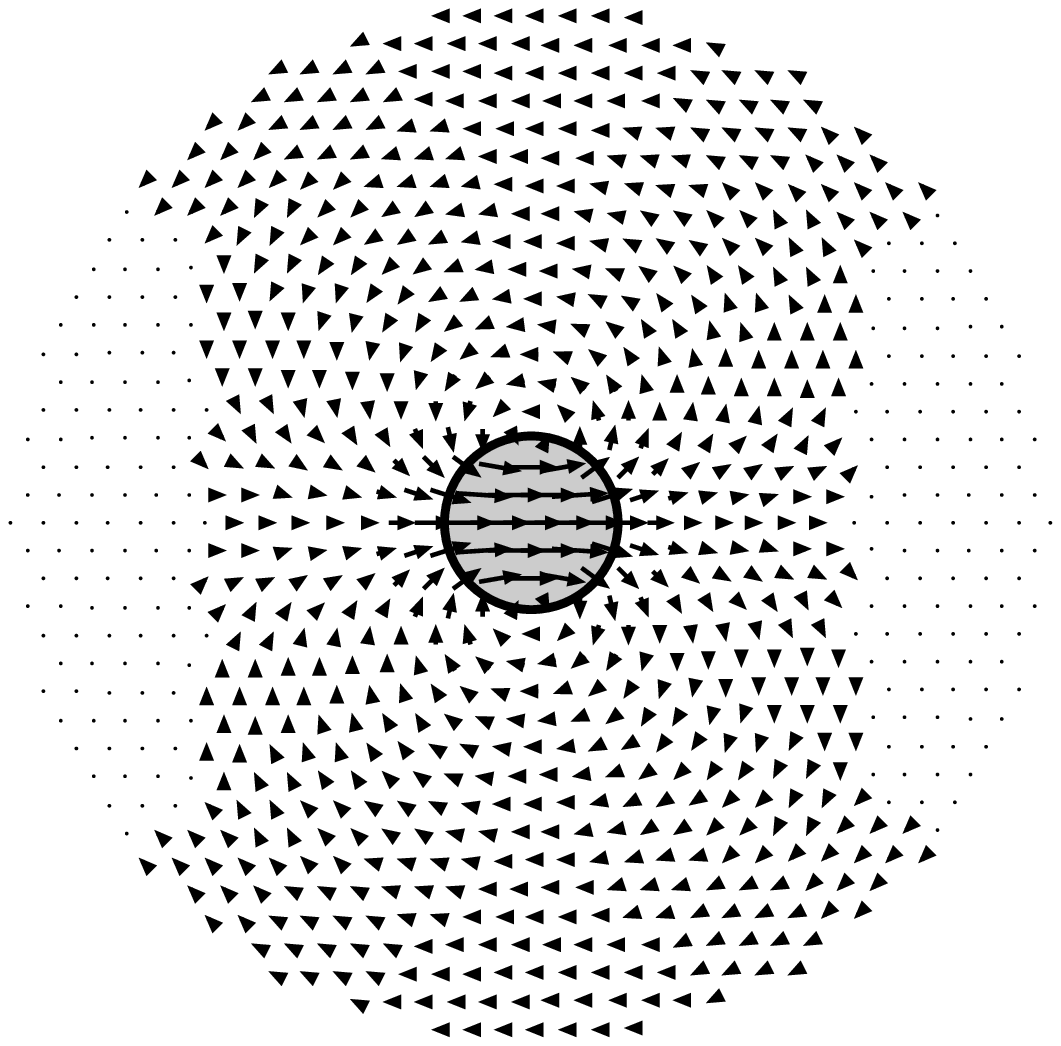}
\end{center}
\caption{ \label{curr}
{\it a.c.} current density pattern for an pinned $s$-wave vortex
with a pinning radius
$\xi_p=1.57 \xi_0$. As pinning site a circular normal conducting
inclusion located at the center of the vortex was assumed.
The border of the pinning center (gray) is shown as black circle.
The frequency of the external electric field with polarization vector
in $x$-direction is chosen
$\omega = 0.1 \Delta $. Left picture for out of phase (reactive) response, 
right picture
for in phase (absorptive) response.
Impurity scattering is taken into account
in Born limit with a mean free path of $\ell=10\xi_0$.
The temperature is $T=0.3T_c$. Calculations are done in the
high-$\kappa $ limit.
}
\end{figure}

\subsection{Induced current density}

Results for the {\it a.c.} component of the current density near a
pinned vortex are shown in Fig. \ref{curr} for $\omega=0.1\Delta$.
In addition to the a.c. current there is the time-independent 
circulating supercurrent around the vortex center which adds 
to the current shown in Fig. \ref{curr}. The current response shows 
a dipolar pattern, which is also observed for unpinned vortices.
The in-phase current response (right panel) indicates a region of strong absorption 
within the pinning region ($\vec{j}||\vE^{\mbox{\tiny ext}}$), and emission
($\vec{j}\cdot\vE^{\mbox{\tiny ext}}<0$) in the region roughly perpendicular 
to the direction of the applied field several coherence lengths away from
the pinning center. Calculations of the energy transport current show that
energy absorbed in the core is transported away from the vortex
center by the vortex core excitations in directions predominantly 
perpendicular to the applied field. The net absorption is ultimately 
determined by inelastic scattering and requires integrating the local 
absorption and emission rate over the vortex array. 
Note that the long-range dipolar component
does not contribute to the total dissipation.
Far from the vortex core the current response is
out of phase with the electric field and 
predominantly a non-dissipative supercurrent.
Also note that at low frequency we clearly 
observe the counterflowing
supercurrent within the pinning center.

\subsection{Summary}

The electrodynamics
of the vortex state in the intermediate-clean regime is nonlocal and
largely determined by the response of the vortex-core states.
Transitions involving the vortex-core states, and their coupling to the 
collective motion of the condensate requires dynamically self-consistent 
calculations of the order parameter, self energies, induced fields, 
excitation spectra and distribution functions. The results of these 
calculations provide new insight into the dynamics of vortex cores 
in conventional and unconventional superconductors.

\subsection{Acknowledgement}

This work was supported in part by the NSF through grant DMR-9972087.


\end{document}